\documentclass[acmsmall,screen=true,anonymous=false,bookmarks=true]{acmart}
\usepackage{amsmath,amsfonts}
\usepackage{upgreek}
\usepackage{needspace}
\usepackage{multirow}
\usepackage{multicol}
\usepackage{amsthm}
\usepackage{listings}
\usepackage{algorithm}
\usepackage{algpseudocode}
\usepackage{graphicx}
\usepackage{textcomp}
\usepackage{xcolor}
\usepackage{cleveref}
\usepackage{booktabs}
\usepackage{comment}
\usepackage{tabularx}
\usepackage{multirow}
\usepackage{bm}
\usepackage{url}
\usepackage{xspace}
\usepackage{pifont}
\usepackage{arydshln}
\usepackage{verbatim}
\usepackage[referable]{threeparttablex}
\usepackage{calc}
\usepackage{soul}
\usepackage{color}
\usepackage{float}
\usepackage{subcaption}
\usepackage[utf8]{inputenc}
\usepackage{booktabs}
\usepackage{tabularx}
\usepackage{tikz}
\usetikzlibrary{shapes.geometric, arrows}

\usepackage{filecontents}
\usepackage{pgfplots}
\usepackage{pgfplotstable}
\usepackage{scalefnt}
\pgfplotsset{compat=newest}
\usepgfplotslibrary{groupplots}

\definecolor{YellowBorder}{RGB}{223,161,89}
\definecolor{YellowFill}{RGB}{254,236,214}
\definecolor{GreenBorder}{RGB}{91,120,60}
\definecolor{GreenFill}{RGB}{243,245,240}
\definecolor{BlueBorder}{RGB}{69,96,160}
\definecolor{PinkBorder}{RGB}{164,42,37}

\newtheorem{theorem}{Theorem}

\usepackage{tcolorbox}
\tcbuselibrary{skins,breakable}

\usepackage{xcolor,graphicx}

\setcopyright{acmcopyright}
\acmJournal{TODAES}
\graphicspath{{./fig/}}

\usepackage{tcolorbox}
\tcbuselibrary{skins,breakable,listings}
\newtcblisting{dslbox}[2][]{listing only, listing engine=listings, listing options={basicstyle=\small\ttfamily, breaklines=true, breakatwhitespace=false, columns=fullflexible, keepspaces=true, showstringspaces=false, upquote=true, extendedchars=true}, colback=YellowFill!50!white,
colframe=YellowBorder!70!white,fonttitle=\scshape\small,
colbacktitle=YellowBorder!70!white,breakable,enhanced,
left=0.5mm,
attach boxed title to top left={xshift=3mm, yshift=-2mm},
title={#2},#1}

\newtcblisting{promptbox}[2][]{listing only, listing engine=listings, listing options={basicstyle=\scriptsize\ttfamily, breaklines=true, breakatwhitespace=false, columns=fullflexible, keepspaces=true, showstringspaces=false, upquote=true, extendedchars=true}, colback=GreenFill!40!white,
colframe=GreenBorder!70!white,fonttitle=\scshape\small,
colbacktitle=GreenBorder!80!white,breakable,enhanced,
left=0.5mm,
attach boxed title to top left={xshift=3mm, yshift=-2mm},
title={#2},#1}

\newtheorem{definition}[theorem]{Definition}

\begin{document}

\setcounter{page}{1}

\title{\textit{PrefixAgent}: An LLM-Powered Design Framework for Efficient Prefix Adder Optimization}

\author{Dongsheng Zuo}
\affiliation{
    \institution{The Hong Kong University of Science and Technology (Guangzhou)}
}
\author{Jiadong Zhu}
\affiliation{
    \institution{The Hong Kong University of Science and Technology (Guangzhou)}
}
\author{Yang Luo}
\affiliation{
    \institution{The Hong Kong University of Science and Technology (Guangzhou)}
}
\author{Yuzhe Ma}
\authornote{Corresponding author: Yuzhe Ma.}
\affiliation{
    \institution{The Hong Kong University of Science and Technology (Guangzhou)}
}

\begin{abstract}
Prefix adders are fundamental arithmetic circuits, but their design space grows exponentially with bit-width, posing significant optimization challenges.
Previous works face limitations in performance, generalization, and scalability.
To address these challenges, we propose \textit{PrefixAgent}, a large language model (LLM)-powered framework that enables efficient prefix adder optimization.
Firstly, we reformulate the problem into two subtasks, namely backbone synthesis and structure refinement, which effectively reduces the search space.
The LLM performs these two phases of optimization by invoking tools through function calls, enabling it to iteratively construct the backbone and refine local structures based on reasoning and EDA feedback.
Secondly, this new design perspective allows us to systematically collect large-scale, high-quality supervision data.
We leverage the rewriting and equality saturation capabilities of e-graphs to comprehensively explore the prefix adder solution space.
In addition, the explainability of e-graphs enables us to extract fine-grained rewrite trajectories, which serve as interpretable and effective training data.
\textit{PrefixAgent} is then fine-tuned on this dataset, significantly enhancing its optimization ability and reasoning generalization.
Experimental results show that \textit{PrefixAgent} synthesizes prefix adders with smaller areas than baseline methods in nearly all configurations, with the advantage growing at larger bit-widths, and it remains effective under a commercial EDA flow.
\end{abstract}



\maketitle

\pagestyle{plain}

\section{Introduction}
\label{sec:intro}

Adders are fundamental arithmetic components used in digital systems like processors, digital signal processors, and AI chips.
Among various adder architectures, the prefix adder is well-known for its parallel structure and efficiency.
The prefix adder's internal operation is represented by a prefix graph, where each node computes group propagate and generates signals for rapid parallel carry computation.
However, the design space of prefix adders grows exponentially with bit-width, making exhaustive exploration impractical.
Traditionally, adder designs rely on regular structures or cumbersome manual optimization.
Therefore, developing efficient and flexible adder design methodologies is essential.

Previous works on adder design primarily involve regular structures and automated design methods.
Regular adder structures such as Kogge–Stone~\cite{Datapath-1973TC-Kogge}, Sklansky~\cite{Datapath-1960TC-Sklansky}, Brent–Kung~\cite{Datapath-1982TC-Brent}, and Han–Carlson~\cite{Datapath-1987ARITH-Han} have been proposed, focusing on trade-offs among logic level, maximum fanout, and wiring tracks.
However, these regular structures may not satisfy diverse requirements.
Practical design requires balancing area, delay, and power while considering timing constraints as well as various input arrival time profiles.
Consequently, various heuristics and search-based methods are proposed~\cite{Datapath-1990DAC-Fishburn,1996-IWLAS-Zimmermann,Datapath-2005ASPDAC-Zhu,Datapath-2024ICCAD-Zuo,Datapath-2003ICCAD-Liu,Datapath-2007GLSVLSI-Matsunaga,Datapath-2024DAC-Lin,Datapath-2013DAC-Roy,Datapath-2014TCAD-Roy,Datapath-2016TACD-Roy}, which improve the design efficiency to some extent.
However, it is difficult to design a generic optimization approach that can cover a rich set of design scenarios.
In addition, a non-trivial gap persists between structural-level metrics and physical-implementation metrics.
Machine learning (ML) methods have been proposed to bridge this gap.
Recent works~\cite{Datapath-2017ISLPED-Roy,DSE-2019TCAD-MA,DSE-2022TCAD-Geng} utilized ML models to predict the metrics and optimize designs over pruned adder solution spaces.
More recently, reinforcement learning (RL)~\cite{RL-2021DAC-Roy,Datapath-2023DAC-Zuo,Datapath-2025TODASE-Zuo,Datapath-2024ICML-Wang,Datapath-2025ICLR-Wang}, variational autoencoder (VAE)~\cite{Datapath-2024DAC-Song}, and Monte Carlo tree search (MCTS)~\cite{Datapath-2024NIPS-Lai} methods have demonstrated progress in arithmetic circuit optimization tasks.
These methods iteratively modify structures, guided by actual physical metrics.
However, these methods suffer from challenges in generalization and scalability due to the extensive training required for each new scenario and the inefficiency of trial-and-error exploration in large action spaces.

\begin{figure}[!tb]
\centering
\includegraphics[width=.8\linewidth]{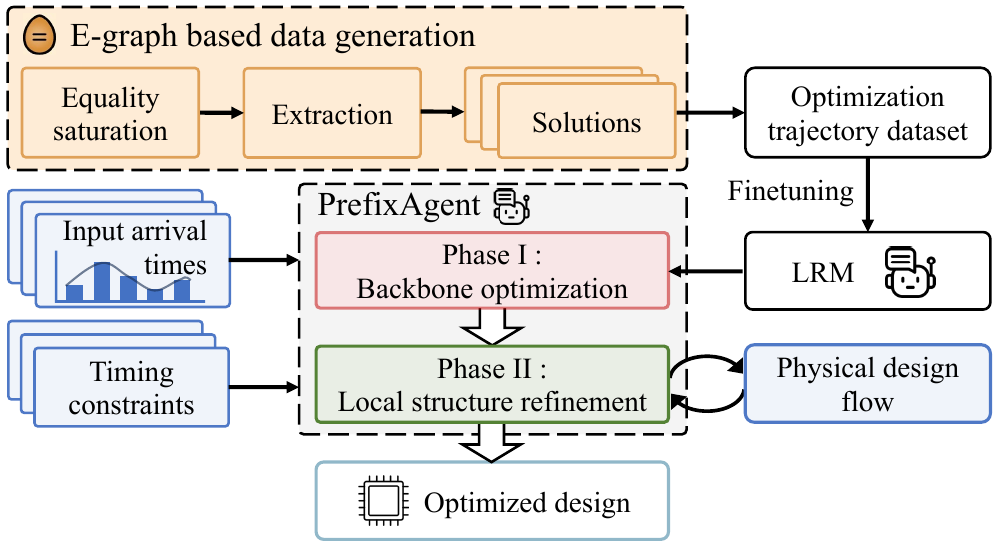}
\caption{The two-phase framework in \textit{PrefixAgent}.}
\label{fig:intro}
\end{figure}

Inspired by manual design practices, experienced engineers typically choose a suitable initial structure based on design specifications and refine it using detailed feedback from EDA tools.
Although effective, this process requires significant engineering effort.
An intelligent agent with strong reasoning and interactive capabilities can automate the design process, addressing generalization challenges and enabling more efficient, scalable optimization.
Motivated by this, emerging large language models (LLMs) provide a promising solution to further automate and optimize the prefix adder design process.
Recent advancements in large reasoning models (LRMs) \cite{openai2024learning,openai2025o3,alibabaQwQMaxPreview, LLM-2025Arxiv-DeepSeek-AI, alibabaQwQ32B} have demonstrated their powerful reasoning capabilities and potential for solving complex problems.
In circuit design tasks, LLMs also have shown potential in analog circuit design~\cite{Analog-2024AAAI-Lai, Analog-2024ICML-Chang, Analog-2025ICLR-Gao} and RTL code design \cite{RTL-2024DAC-Chang}.
However, applying LRMs to the prefix adder design task still faces two main challenges.
The first challenge is the limited capability of large models to efficiently generate and reason about complex graphs with more than 20 nodes \cite{LLM-2024KDD-Chen, LLM-2025ILCR-Dai, LLM-2025ILCR-Tang}.
This issue is particularly evident in the EDA domain, where existing LLM-based methods have shown promise for small graph circuits, such as analog circuits with fewer than 10 nodes~\cite{Analog-2024AAAI-Lai, Analog-2024ICML-Chang}.
Recent work \cite{Datapath-2024Arxiv-Xiao} attempts to employ LRMs to generate textual representations of prefix graphs; however, it still faces graph reasoning challenges.
Specifically, the node count in a prefix graph theoretically scales quadratically with bit-width, and these graphs must adhere to stringent structural constraints, further complicating the generation process.
The second challenge is the lack of high-quality data, which is essential for fine-tuning LRMs on the prefix adder design task.
Recent studies highlight the importance of high-quality data in improving the reasoning capabilities of LRMs \cite{LLM-2025Arvix-Ye, LLM-2025Arvix-Li, LLM-2025Arvix-Muennighoff}.
However, simple random data generation or perturbation-based strategies are ineffective in producing meaningful optimization trajectories for training.

In this paper, we propose \textit{PrefixAgent}, an LLM-powered framework that enables efficient prefix adder optimization.
The overall workflow of the framework is illustrated in \Cref{fig:intro}.
To address LLMs' limitations on complex graphs, our framework considers the prefix adder design problem from a new perspective, which decomposes the prefix adder synthesis task into two subtasks, namely \textit{backbone generation} and \textit{local structure refinement}.
The backbone, representing the critical structural component of the adder, significantly simplifies the overall design complexity.
The new perspective also enables the LLM to efficiently generate zero-deficiency~\cite{Datapath-1986JA-Snir} or low-deficiency adders.
Rather than requiring the LLM to generate fully valid prefix structures with high complexity, \textit{PrefixAgent} adopts the tool-integrated reasoning (TIR) technique~\cite{gou2023tora,li2025torl}, which delegates precise local modifications to external tools during the reasoning process.
This empowers the LLM's reasoning capabilities, allowing it to focus on high-level decision-making based on detailed feedback after physical implementation.
Furthermore, based on the backbone-centric optimization paradigm for prefix adders, we observed a correspondence between backbone optimization and e-graph-based rewriting.
This enables us to gather high-quality data and interpretable reasoning trajectories using e-graphs, successfully addressing the second challenge, i.e., data scarcity, in building LLM agents.
By fine-tuning the LLM on these carefully curated datasets, \textit{PrefixAgent} learns to efficiently optimize adders under various scenarios and design requirements.
Compared to RL~\cite{RL-2021DAC-Roy}, MCTS~\cite{Datapath-2024NIPS-Lai}, VAE~\cite{Datapath-2024DAC-Song}, PrefixGPT~\cite{Datapath-2026AAAI-Ding}, and prior LLM methods~\cite{Datapath-2024Arxiv-Xiao}, \textit{PrefixAgent} generates prefix adders with smaller areas in nearly all cases, with a widening advantage at larger bit-widths.
Under commercial physical design flows, it achieves superior results to adders generated by commercial tools.
The key contributions of this work are summarized as follows:
\begin{itemize}
    \item We propose \textit{PrefixAgent}, an LLM-powered framework for efficient prefix adder optimization, representing the first LLM-based approach that scales to 64-bit designs and outperforms commercial tools.
    \item A new design perspective is proposed for the prefix adder, which consists of backbone generation and local refinement. An agent is utilized to perform reasoning for efficient backbone synthesis and targeted optimization guided by detailed EDA feedback.
    \item Owing to the new design perspective, we utilize the e-graph-based method to systematically generate comprehensive, interpretable, and high-quality optimization trajectories for the backbone optimization task, enabling effective fine-tuning of the agent.
    \item Experimental results show that \textit{PrefixAgent} synthesizes prefix adders with smaller areas than baseline methods in nearly all configurations, and outperforms the commercial synthesis tool.
\end{itemize}

\section{Preliminary}
\label{sec:prelim}

\begin{figure}[!tb]
\centering
\includegraphics[width=0.5\linewidth]{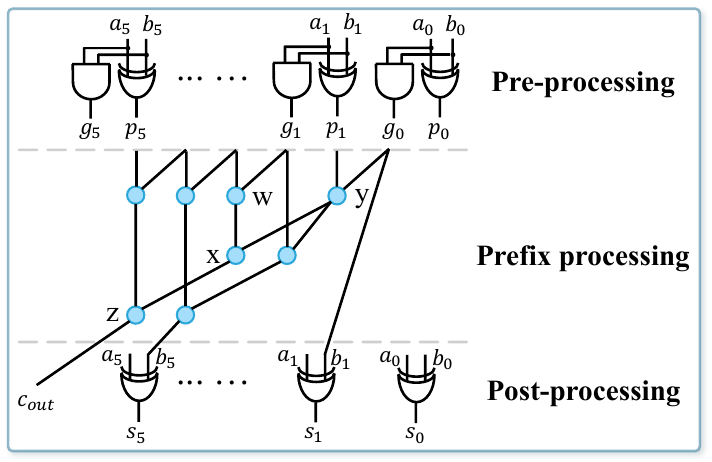}
\caption{Structure of a prefix adder.}
\label{fig:prelim-adder}
\end{figure}

\subsection{Prefix Adder}
\label{sec:prelim:prefix}
Given two $n$-bit binary numbers, $A = a_{n-1}a_{n-2}\dots a_0$ and $B = b_{n-1}b_{n-2}\dots b_0$, the prefix adder computes the sum $S = s_{n-1}s_{n-2}\dots s_0$ and the carry-out bit $C_{out}$.
The operation is performed in three stages: pre-processing, prefix computation, and post-processing.

\textbf{Pre-processing:} Each bit position independently computes propagate ($p_i$) and generate ($g_i$) signals:
\begin{equation}
p_i = a_i \oplus b_i, \quad g_i = a_i \cdot b_i.
\end{equation}

\textbf{Prefix Computation:} The propagate and generate signals are combined recursively using the associative prefix operator \(\mathit{o}\) defined as:
\begin{align}
(G, P)_{i:j} &= (G, P)_{i:k} \;\mathit{o}\; (G, P)_{k-1:j} \notag \\
&= \Big(G_{i:k} + P_{i:k} \cdot G_{k-1:j},\; P_{i:k} \cdot P_{k-1:j}\Big).
\end{align}
As the prefix operator \(\mathit{o}\) is associative, it allows flexibility in grouping strategies.
Each \(\mathit{o}\) operation corresponds to a node in the prefix graph, and different groupings lead to different graph topologies.
This recursive process continues until each bit position \(i\) is assigned an output node to compute \(G_{i:0}\), forming the final prefix structure.

\textbf{Post-processing:} The final carry and sum bits are computed using the prefix results:
\begin{equation}
c_i = G_{i:0}, \quad s_i = p_i \oplus c_{i-1}
\end{equation}

We adopt the notation from~\cite{Datapath-2014TCAD-Roy}, where each prefix node \((i,j)\) represents the computation of \((i,k)\) and \((k\!-\!1,j)\), with \(i \geq k > j\).
Here, \(i\) and \(j\) denote the most significant bit (MSB) and least significant bit (LSB), respectively.
We refer to \((i,k)\) as the \textit{trivial fanin}(\(fi_{t}\)) and \((k\!-\!1,j)\) as the \textit{non-trivial fanin}(\(fi_{nt}\)) of node \((i,j)\).
We further classify the fanouts of each node into \textit{trivial fanout}(\(fo_{t}\)) and \textit{non-trivial fanout}(\(fo_{nt}\)):
A fanout is labeled as \(fo_{t}\) if it shares the same MSB as its parent, and \(fo_{nt}\) if the MSB is larger.
For example, node \((i,j)\) is a \(fo_{t}\) of \((i,k)\) and a \(fo_{nt}\) of \((k\!-\!1,j)\).
In \Cref{fig:prelim-adder}, node \(x\) is a \(fo_{t}\) of node \(w\) and a \(fo_{nt}\) of node \(y\).

Under the standard $(i,j)$ grid representation, an $n$-bit prefix graph has at most $O(n^2)$ potential node positions.
Given a nodelist, we first map it to this grid and, for each existing node, verify the consistency of its fanin (and optionally fanout) relations in constant time, leading to an $O(n^2)$ worst-case complexity for checking whether the prefix graph is legal.

\begin{figure}[!tb]
\begin{minipage}{.48\textwidth}
    \centering
    \includegraphics[width=.9\linewidth]{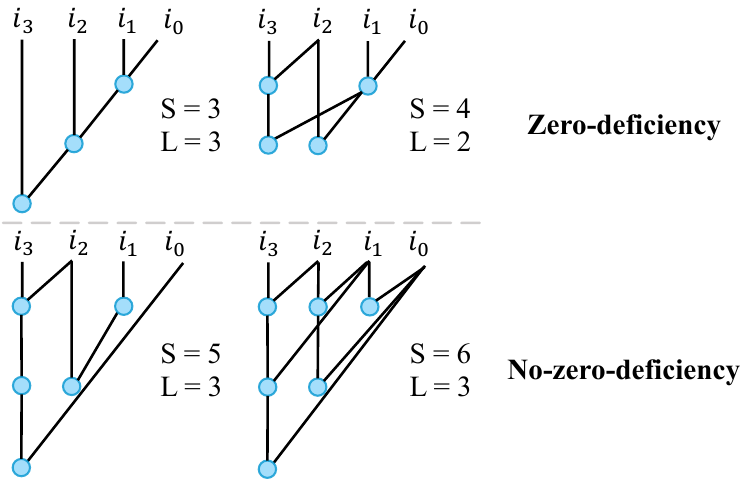}
    \caption{Comparison of zero and non-zero-deficiency adders.}
    \label{fig:prelim-zero}
\end{minipage}
\begin{minipage}{.48\textwidth}
    \centering
    \includegraphics[width=.9\linewidth]{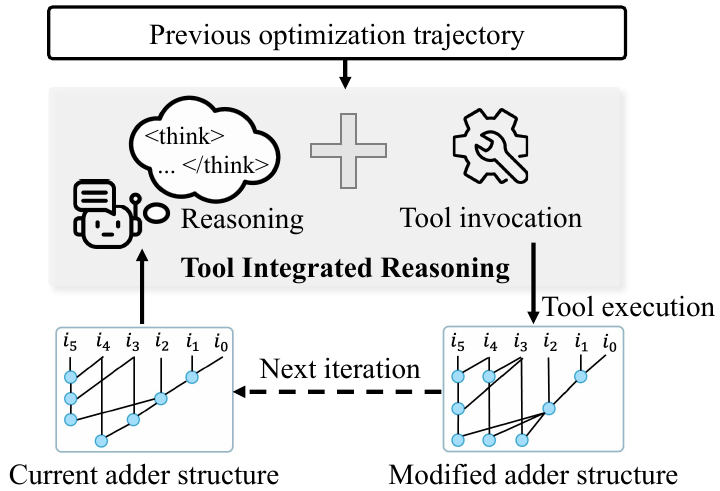}
    \caption{Tool integrated reasoning in \textit{PrefixAgent}}
    \label{fig:tir}
\end{minipage}
\end{figure}

A zero-deficiency adder aims for a balanced design with minimal cost in both prefix graph size and depth.
According to Snir's theorem \cite{Datapath-1986JA-Snir}, a prefix adder is defined as zero-deficiency if its size and depth satisfy the equation:
\begin{equation}
  S + L = 2n - 2,
  \label{eq:zero-deficiency}
\end{equation}
where $S$ and $L$ represent the size and logic level of an $n$-bit prefix adder, respectively.
Examples of zero and non-zero deficiency adders are shown in \Cref{fig:prelim-zero}.

\subsection{Reasoning Models and Function Calling}

Recent large language models (LLMs) such as OpenAI's o1 \cite{openai2024learning} and o3 \cite{openai2025o3}, DeepSeek-R1 \cite{LLM-2025Arxiv-DeepSeek-AI}, and QwQ \cite{alibabaQwQMaxPreview,alibabaQwQ32B} have advanced significantly in reasoning capabilities.
By incorporating intermediate “thoughts”, enclosed within \texttt{<think>...</think>} tags that represent human-like reasoning steps, LLMs are being transformed from simple autoregressive conversational models into more sophisticated cognitive reasoners, referred to as large reasoning models (LRMs).
Specifically, LRMs are encouraged to generate deliberate reasoning steps before generating final responses through post-training strategies such as \cite{shao2024deepseekmath}.
In addition, supervised fine-tuning (SFT) of reasoning models relies on chain-of-thought (CoT) reasoning formatted within \texttt{<think>...</think><answer>...</answer>} tags to ensure consistency and effectiveness.
LLMs have implicitly learned domain-specific circuit design knowledge during pretraining and fine-tuning, and LRMs can generalize this knowledge through structured reasoning to handle complex prefix adder synthesis tasks.

\subsection{Tool Integrated Reasoning}
\label{subsec:tir}

Function calling enables LLMs to delegate precise tasks to external computational tools through structured calls.
In \textit{PrefixAgent}, function calling allows the LLM to invoke external tools for accurate local modifications and evaluations, effectively bridging high-level reasoning with algorithmic precision.

This capability forms the foundation of the \emph{tool integrated reasoning} (TIR) paradigm, where reasoning and tool usage are interleaved throughout the problem-solving process. In a typical TIR loop, the LLM performs symbolic reasoning, emits tool calls based on intermediate conclusions, receives feedback from the tool execution, and updates its internal state accordingly. This paradigm has been shown effective in various domains such as mathematical problem solving, program analysis, and circuit optimization, where purely language-based reasoning often falls short in terms of factual fidelity or structural validity.

As shown in \Cref{fig:tir}, \textit{PrefixAgent} adopts this reasoning paradigm by integrating structured function calling into both backbone optimization and local refinement stages.
To perform transformations on prefix structures, the LLM invokes function calls, each of which is accompanied by tool-generated feedback to guide its subsequent reasoning.

\subsection{E-graphs}

\begin{figure*}[!t]
  \centering
\includegraphics[width=0.95\linewidth]{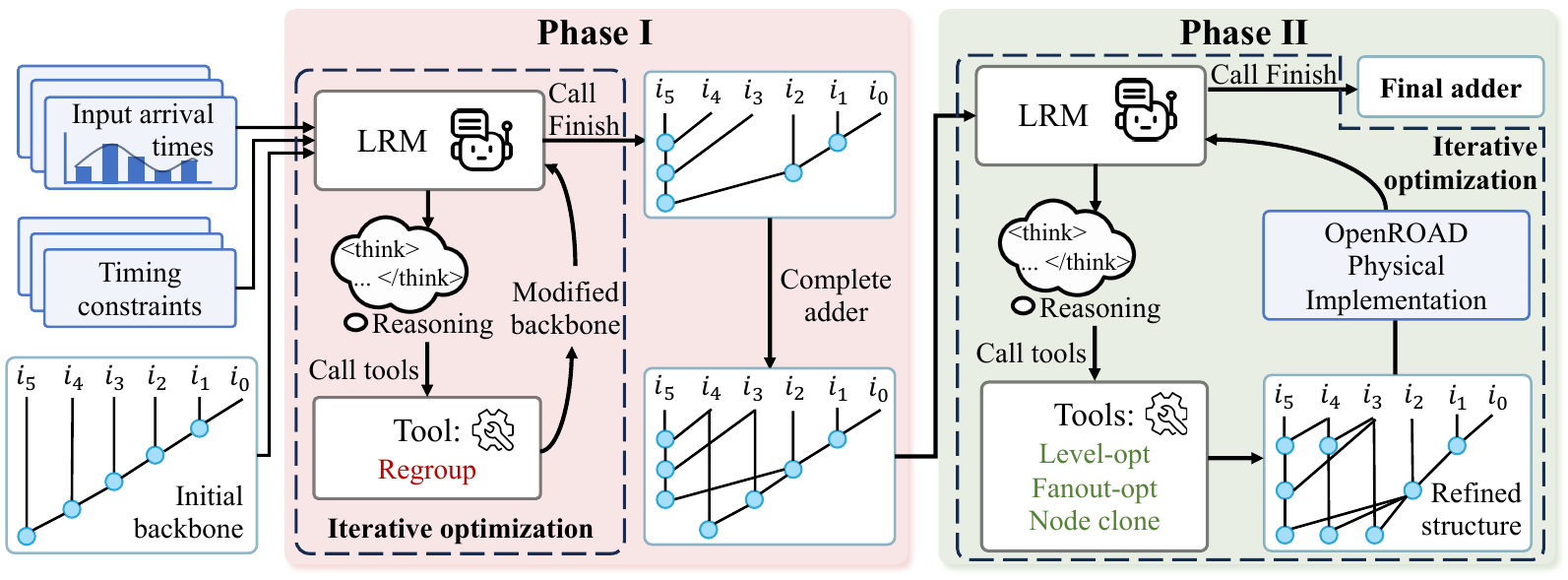}
  \caption{\textit{PrefixAgent} framework overview.}
  \label{fig:method:overview}
\end{figure*}

E-graphs~\cite{Egraph-1980PHD-Nelson} are data structures originally developed in the theorem proving community to compactly represent equivalence classes of expressions.
An e-graph encodes a set of expressions by grouping nodes—representing variables, constants, or operators—into equivalence classes called \textit{e-classes}. Each e-class contains one or more \textit{e-nodes}, which are individual operator applications with pointers to their child e-classes.
In this way, a small number of nodes can represent exponentially many equivalent expressions.
E-graphs grow through a process known as \textit{equality saturation}~\cite{Egraph-2009POPL-Tate}, where all applicable rewrites are repeatedly and non-destructively applied to the structure.
This process guarantees that rewrites do not interfere with each other, avoiding the classical \textit{phase-ordering problem} in compiler and synthesis optimization.
Specifically, rewrites in e-graphs are \textit{constructive}: when a rewrite rule $(\textit{lhs}, \textit{rhs})$ matches an expression in the e-graph, the right-hand side is simply added to the same e-class without removing the left-hand side.
As a result, the e-graph grows monotonically, and all discovered equivalences are preserved throughout the saturation process. The rewrite process eventually reaches a fixed point called \textit{saturation}, where no further rewrites can be applied.

A widely used implementation of e-graphs is the \texttt{egg} library~\cite{Egraph=2021PL-Willsey}, which has enabled extensive research in program optimization and hardware synthesis. In addition to core equality saturation, \texttt{egg} supports several advanced features. Notably, its \textit{e-class analysis} framework allows users to attach semantic attributes to e-classes, enabling cost-guided optimization and program analysis.

E-graphs have been widely adopted in various domains beyond their original use in theorem proving.
They have played a critical role in compiler infrastructures for program optimization, where equality saturation enables robust and phase-ordering-free transformations.
More recently, e-graphs have gained traction in the EDA community.
Specifically, they have been used for RTL-level datapath optimization and equivalence checking of complex arithmetic circuits~\cite{Egraph=2022FMCAD-Coward,Egraph-2024TCAD-Coward}.
For example, the IMpress framework~\cite{Egraph=2022FCCM-Ustun} employs e-graphs to explore alternative decomposition strategies for large integer multipliers in high-level synthesis (HLS).
At the bit-level, E-Syn~\cite{Egraph-2024DAC-Chen} applies e-graph-based rewriting for logic-level optimization.

In this work, we leverage the powerful rewriting capabilities of e-graphs in combination with multiple extraction strategies to generate optimized datasets for the backbone of prefix adders.
Specifically, we utilize equality saturation and custom rewrite rules to explore the design space, and apply timing-aware extraction to identify efficient backbone candidates.
Furthermore, we exploit the \texttt{egg} library's explanation mechanism to extract detailed, interpretable rewrite trajectories, which are then used to construct high-quality supervision data for fine-tuning the LRM.

\section{Proposed PrefixAgent Framework}
\label{sec:algo}
\subsection{Overview}

An overview of the \textit{PrefixAgent} framework is shown in \Cref{fig:method:overview}.
The \textit{PrefixAgent} framework operates in two phases.
In Phase I, a fine-tuned LRM iteratively refines an initial prefix adder backbone structure to achieve an optimized backbone.
Upon completion of Phase I, the framework constructs a full prefix adder based on the optimized backbone, transitioning into Phase II.
In Phase II, the LRM further performs local structural refinement, guided by feedback from the physical implementation to optimize timing performance.
By decomposing the complex adder design task into these two phases, we simplify the complex prefix design task.
The backbone refinement process directly corresponds to e-graph rewrites, enabling efficient generation of high-quality optimization data for fine-tuning the LRM.
The detailed methodology for generating data utilizing e-graphs will be elucidated in \Cref{sec:egraph}.

\textit{PrefixAgent} is a tool integrated reasoning style framework.
Rather than directly generating a fully valid prefix adder structure, \textit{PrefixAgent} employs function calling during reasoning, allowing the LRM to focus more on high-level decision-making, such as where and what type of optimization to perform.
The actual execution and evaluation of structural modifications is handled by external tools.
The LRM determines the end of each phase by explicitly invoking the corresponding completion tools.
The tools employed within each phase are summarized in \Cref{tab:tools}.

\begin{table}[!tb]
\centering
\caption{Summary of tool calls in \textit{PrefixAgent}}
\label{tab:tools}

\begin{tabular}{l|ll}
\hline
                          & Tool & Description \\
                          \hline
\multirow{2}{*}{Phase~ I}  & \texttt{regroup} & Modify backbone by regroup 2 nodes.            \\

                          & \texttt{finish 1}     &        Complete Phase~I.     \\
                          \hline
\multirow{4}{*}{Phase~ II} & \texttt{level-opt}     &    Reduce logic level of target node.       \\
                          & \texttt{fanout-opt}     &   Reduce fanout of target node.          \\
                          & \texttt{node clone}    &    Clone high fanout node.         \\
                          & \texttt{finish 2}     &    Complete Phase~II.     \\
                          \hline
\end{tabular}

\end{table}

\begin{figure}[!tb]
\centering
\includegraphics[width=.65\linewidth]{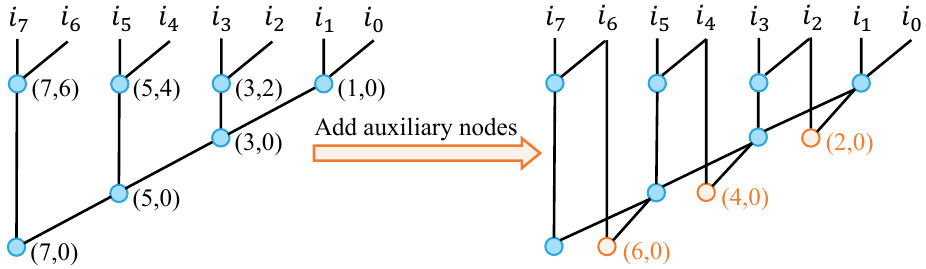}
\caption{Illustration of backbone (left) and completed prefix adder (right) with added auxiliary nodes.}
\label{fig:method:backbone}
\end{figure}

\begin{figure}[!tb]
\centering
\includegraphics[width=0.65\linewidth]{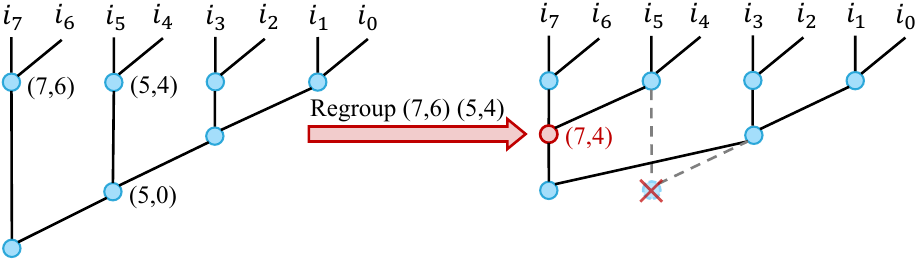}
\caption{Backbone regroup example.}
\label{fig:method:regroup}
\end{figure}

\subsection{Backbone of Prefix Adder}
\label{subsec:backbone}

\begin{definition}[Backbone]
Given an \(N\)-bit prefix graph, the backbone is defined as the subgraph consisting of nodes that compute the carry for the most significant bit (MSB), denoted as node \((N-1,0)\).
\end{definition}

As illustrated in the left part of \Cref{fig:method:backbone}, the backbone for an 8-bit prefix adder includes nodes specifically structured for computing the MSB carry. The backbone dominates the fundamental architecture of the prefix adder, significantly influencing its overall performance and complexity.
In this example, the backbone corresponds to the following grouping structure:
\begin{equation}
G_{7:0} = (i_0 \,\mathit{o}\, i_1) \,\mathit{o}\, (i_2 \,\mathit{o}\, i_3) \,\mathit{o}\, (i_4 \,\mathit{o}\, i_5) \,\mathit{o}\, (i_6 \,\mathit{o}\, i_7)
\label{eq:backbone1}
\end{equation}

The backbone can be viewed as a full binary tree with $N$ leaves, where pairs of input nodes from \(i_0\) to \(i_{N-1}\) are recursively grouped.
It is well known that such a binary tree has exactly $N-1$ nodes.
Therefore, we let $S_B$ denote the number of backbone nodes, and $S_B = N - 1$.
To derive the full prefix adder from a given backbone, each bit position must be associated with an output node.
For a backbone with level \(L_B\), there already exist \(L_B\) output nodes corresponding to the backbone level.
Therefore, for the remaining \(N - 1 - L_B\) bit positions that lack outputs, we insert auxiliary nodes to complete the prefix graph as shown in \Cref{fig:method:backbone}.
The total number of auxiliary nodes is thus \(S_A = N - 1 - L_B\).
This construction ensures that all intermediate nodes have non‑trivial fanout, i.e., each intermediate propagate–generate signal is utilized by subsequent nodes.
According to \cite{Datapath-1986JA-Snir}, this is a necessary condition for a zero-deficiency adder.
By appropriately designing the backbone structure, the completed prefix adder can achieve a level \(L\) equal to the backbone level \(L_B\), and the total size and level jointly satisfy \Cref{eq:zero-deficiency}, corresponding to a zero-deficiency adder.
In scenarios involving non-uniform input arrival times, the final level \(L\) may exceed \(L_B\).
In such cases, subsequent local refinements using the \texttt{level opt} tool in Phase~ II can yield a low-deficiency adder.

Focusing on the backbone reduces the design complexity while resulting in zero- or low-deficiency adders.
For an \(N\)-bit prefix adder, the full solution space grows exponentially as a product of Catalan numbers.
Restricting optimization to the backbone reduces this to a single Catalan number.
For example, in the 16-bit case, the design space drops from approximately \(10^{48}\) to \(10^{6}\).

\subsection{Regroup of Backbone}
\label{subsec:backbone_regroup}

To further improve the backbone structure, \textit{PrefixAgent} employs a regrouping strategy that iteratively modifies the grouping pattern within the binary tree.
A regroup operation alters the backbone by reorganizing two existing subtrees.
As illustrated in \Cref{fig:method:regroup}, regrouping nodes \((7,6)\) and \((5,4)\) introduces a new node \((7,4)\) and removes node \((5,0)\), effectively restructuring the backbone.
In this example, the grouping structure of the backbone changes from \Cref{eq:backbone1} to:
\begin{equation}
G_{7:0} = ((i_0 \,\mathit{o}\, i_1) \,\mathit{o}\, (i_2 \,\mathit{o}\, i_3)) \,\mathit{o}\, ((i_4 \,\mathit{o}\, i_5) \,\mathit{o}\, (i_6 \,\mathit{o}\, i_7))
\label{eq:backbone2}
\end{equation}
The detailed implementation of regroup operations, including how regroup candidates are identified within a backbone, is presented in \Cref{subsec:phase1}.
To identify valid regrouping opportunities, we impose structural constraints.
Specifically, two candidate nodes \(a\) and \(b\) must satisfy the condition \(a.\mathit{lsb} = b.\mathit{msb} - 1\), ensuring they can serve as valid non-trivial and trivial fanins for a new node.
Candidate pairs are enumerated by traversing the current set of \(N-1\) backbone nodes, and output nodes are excluded from regrouping.

Notably, each regroup operation can be precisely interpreted as an expression-level rewrite of the MSB carry computation, for example, transforming the expression in \Cref{eq:backbone1} into that in \Cref{eq:backbone2}.
This aligns with the equivalence-preserving rewrites in e-graphs.
This elegant correspondence enables us to leverage e-graph to generate high-quality backbone structures along with their corresponding regrouping trajectories, which can be used to construct effective fine-tuning data for the LRM, as described in detail in \Cref{sec:egraph}.

\subsection{PrefixAgent Phase~ I: Backbone}
\label{subsec:phase1}

\begin{algorithm}[t]
  \caption{Backbone Optimization by LRM}
  \label{alg:backbone_short}
  \begin{algorithmic}[1]
    \Require Bit-width $N$, arrival time $\mathcal{D}$, target delay $T$, max iteration $K$
    \Ensure Optimized backbone $B^\star$
    \State $B \gets \text{InitBackbone}(N)$ \Comment{Initialize serial backbone} \label{alg:1:1}
    \For{$k = 1$ to $K$}  \label{alg:1:2}
      \State $\mathcal{C} \gets$ FindCandidates($B$) \Comment{Get regroup candidates}\label{alg:1:3}
      \State prompt $\gets$ BuildPrompt($B$, $\mathcal{C}$, $\mathcal{D}$, $T$)  \label{alg:1:4}
      \State call $\gets$ LRM(prompt) \Comment{LRM decision} \label{alg:1:5}
      \If{call == \texttt{finish}}  \Comment{Terminate by LRM} \label{alg:1:6}
        \State break  \label{alg:1:7}
      \Else \label{alg:1:8}
        \State $(a, b) \gets$ call \label{alg:1:9}
        \State $B \gets$ Regroup($a$, $b$) \Comment{Apply regroup} \label{alg:1:10}
      \EndIf \label{alg:1:11}
    \EndFor \label{alg:1:12}
    \State \Return $B$ \Comment{Return final backbone} \label{alg:1:13}
    \Function{FindCandidates}{$B$} \label{alg:1:14}
      \State $\mathcal{P} \gets \emptyset$ \label{alg:1:15}
      \For{$i = N-1$ to $1$}  \label{alg:1:16}
        \State $a \gets$ node in column $i$ with lsb $> 0$ \label{alg:1:17}
        \State $b \gets$ node in column $a.\textit{lsb} - 1$ with lsb $> 0$  \label{alg:1:18}
        \State $\mathcal{P}.\text{append}((a, b))$  \label{alg:1:19}
      \EndFor \label{alg:1:20}
      \State \Return $\mathcal{P}$ \label{alg:1:21}
    \EndFunction \label{alg:1:22}
    \Function{Regroup}{$a, b$} \label{alg:1:23}
      \State Add node $(a.\textit{msb}, b.\textit{lsb})$ \label{alg:1:24}
      \State Remove node $(b.\textit{msb}, 0)$ \label{alg:1:25}
    \EndFunction \label{alg:1:26}
  \end{algorithmic}
\end{algorithm}

As shown in \Cref{fig:method:overview}, given the input arrival time profile and the timing constraint specified by the user, the first phase of \textit{PrefixAgent} focuses on backbone generation.
The overall optimization process is outlined in Algorithm~\ref{alg:backbone_short}.
The framework first initializes a fully serial backbone structure \(B\), which serves as the starting point for optimization (\Cref{alg:1:1}).
Next, regroup candidates are generated (\Cref{alg:1:3}) by traversing the backbone and identifying structurally valid node pairs \((a, b)\) that satisfy regrouping constraints (\Cref{alg:1:14,alg:1:15,alg:1:16,alg:1:17,alg:1:18,alg:1:19,alg:1:20,alg:1:21,alg:1:22}).
The current backbone representation, together with candidate pairs, are then embedded into a prompt (\Cref{alg:1:4}) and passed to the LRM for decision-making (\Cref{alg:1:5}).
The LRM may respond with a function call to apply a regroup operation (\Cref{alg:1:9}) or decide to terminate the optimization process (\Cref{alg:1:6}).
Specifically, as detailed in \Cref{alg:1:23,alg:1:24,alg:1:25,alg:1:26}, the regroup operation introduces a new node to replace one previous node, thereby modifying the grouping pattern within the backbone.
This iterative procedure continues until either a termination is triggered by the LRM or the maximum number of iterations \(K\) is reached.
The resulting backbone is then used to generate a complete prefix adder structure and passed to the next phase.

To enable the LRM to make effective optimization decisions based on both input arrival times and the current structure, we adopt a Timing-Annotated S-Expression format that annotates each backbone node with its signal arrival time.
To obtain accurate arrival time annotations, we construct the backbone netlist using interleaved \texttt{AOI/OAI} gates and perform static timing analysis (STA).
This process captures accurate signal arrival times for all nodes, including input nodes.
An example Timing-Annotated S-Expression for a fully parallel 4-bit backbone is shown below:

\begin{promptbox}{Timing-Annotated S-Expression}
(3,0) [arrival=0.1050]
group(
  (3,2) [arrival=0.0600] =
  group(
    input (3,3) [arrival=0.0150],
    input (2,2) [arrival=0.0250]
  ),
  (1,0) [arrival=0.0700] =
  group(
    input (1,1) [arrival=0.0350],
    input (0,0) [arrival=0.0350]))
\end{promptbox}
This representation leverages the inherent hierarchical nature of binary trees and captures the structural information of the backbone clearly.
It also allows the LRM to analyze timing behavior along different propagation paths, thereby supporting more informed reasoning.

\subsection{PrefixAgent Phase~ II: Local Structural Refinement}
\label{subsec:phase2}

After the backbone is optimized and the prefix adder is completed by adding auxiliary nodes, timing violations may still occur.
These violations are primarily caused by three factors:
(1) the overall level \(L\) of the completed prefix adder may exceed the backbone level \(L_B\);
(2) the addition of auxiliary nodes may increase fan-out at certain nodes; and
(3) layout-related effects introduced during physical implementation.

\begin{figure}[!tb]
\centering
\includegraphics[width=.7\linewidth]{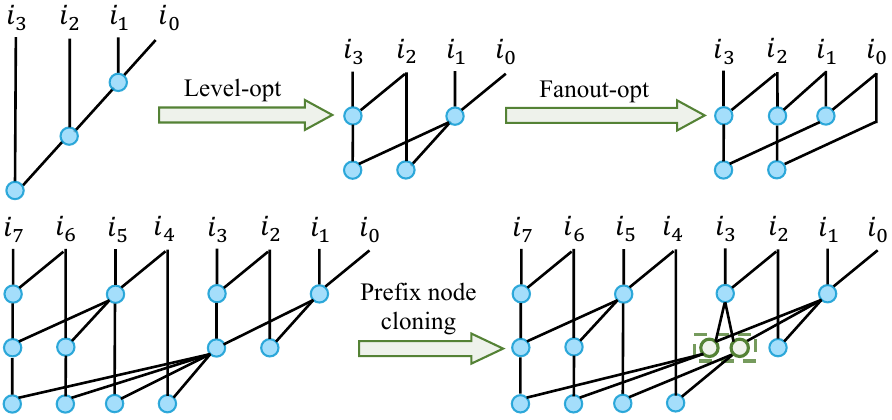}
\caption{Illustration of timing optimization tools: level-opt, fanout-opt and node clone.}
\label{fig:method:timing-tool}
\end{figure}

To resolve these timing violations, \textit{PrefixAgent} enters Phase~II, which applies local structural refinements to improve timing.
In this phase, the LRM identifies timing-critical nodes and invokes one of the three timing optimization tools.
The LRM determines the appropriate optimization target and tool type, while the execution of structural modifications is delegated to external tools.
As illustrated in \Cref{fig:method:timing-tool}, the three tools include:

(1) \texttt{level-opt}, which reduces the target node's level by inserting an additional node;

(2) \texttt{fanout-opt}, which inserts a node to alleviate excessive fanout at the target node;

(3) \texttt{node clone}, which creates a copy of a high-fanout node to share the load.

In our framework, this structural node cloning is complementary to physically aware cloning in downstream implementation tools: cloned nodes can improve placement and routing friendliness as reported in~\cite{Datapath-2016TACD-Roy}, and the node-clone tool can be disabled in \textit{PrefixAgent} if desired.
After each optimization step, we invoke a prefix-graph validator to check structural legality using the $O(n^2)$ procedure described in \Cref{sec:prelim:prefix}; for bit-widths up to 64 bits, this runs at the millisecond scale and does not dominate the overall runtime.

To support effective reasoning during structural refinement, we introduce the \textit{Enhanced Prefix Representation} (EPR).
Preliminary experiments show that explicitly encoding node-level attributes—such as logic level, parent relationships, and fanout—significantly improves the LRM's ability to analyze and optimize prefix graph structures.
The EPR compactly captures both global properties and detailed per-node information of a prefix adder.
It includes input and non-input node lists, level and connection relationships.
An example EPR for a 4-bit prefix adder is shown below:

\begin{promptbox}{Enhanced Prefix Representation (EPR)}
Bitwidth: 4
Non-input nodes: 3
Max level: 3
Max fanout: 1

Input nodes:
(0,0), fo_t: [], fo_nt: [(1,0)] 
(1,1), fo_t: [(1,0)], fo_nt: [] 
(2,2), fo_t: [(2,0)], fo_nt: [] 
(3,3), fo_t: [(3,0)], fo_nt: []

Non-input nodes:
(3,0), lvl: 3, fi_t:(3,3), fi_nt:(2,0), fo_t: [], fo_nt: [] 
(2,0), lvl: 2, fi_t:(2,2), fi_nt:(1,0), fo_t: [], fo_nt: [(3,0)] 
(1,0), lvl: 1, fi_t:(1,1), fi_nt:(0,0), fo_t: [], fo_nt: [(2,0)] 
\end{promptbox}

During timing refinement, critical path information is essential for guiding optimization decisions.
Based on the start and end points of the critical path, we reconstruct the full sequence of nodes along the critical path and generate a structured analysis.
An example for critical path analysis is as follows:

\begin{promptbox}{Critical Path Analysis}
Lvl 0: (0,0), [INPUT] fo_t: [], fo_nt: [(1,0)]
Lvl 1: (1,0), fi_t: (1,1), fi_nt: (0,0), fo_t: [], fo_nt: [(2,0)]
Lvl 2: (2,0), fi_t: (2,2), fi_nt: (1,0), fo_t: [], fo_nt: [(3,0)]
Lvl 3: (3,0), fi_t: (3,3), fi_nt: (2,0), fo_t: [], fo_nt: []

- Level efficiency: 2/3 (theoretical min: 2, actual: 3)
\end{promptbox}

In each refinement iteration, the \textit{PrefixAgent} framework constructs a prompt using current physical metrics, tool feedback, EPR, and the critical path analysis.
In Phase~I, the LRM operates on timing-annotated s-expressions of the MSB-carry backbone, which provides a compact tree topology and path delays.
In Phase~II, it relies on the EPR format to expose full prefix-graph connectivity and node-level attributes, so each phase uses the representation that best matches its optimization objective.
Based on this information, the LRM performs reasoning and then either invokes an appropriate optimization tool or issues a termination call to finalize the design and prevent over-design.

\section{E-graph-based Data Generation}
\label{sec:egraph}

\begin{figure*}[!t]
  \centering
\includegraphics[width=1.\linewidth]{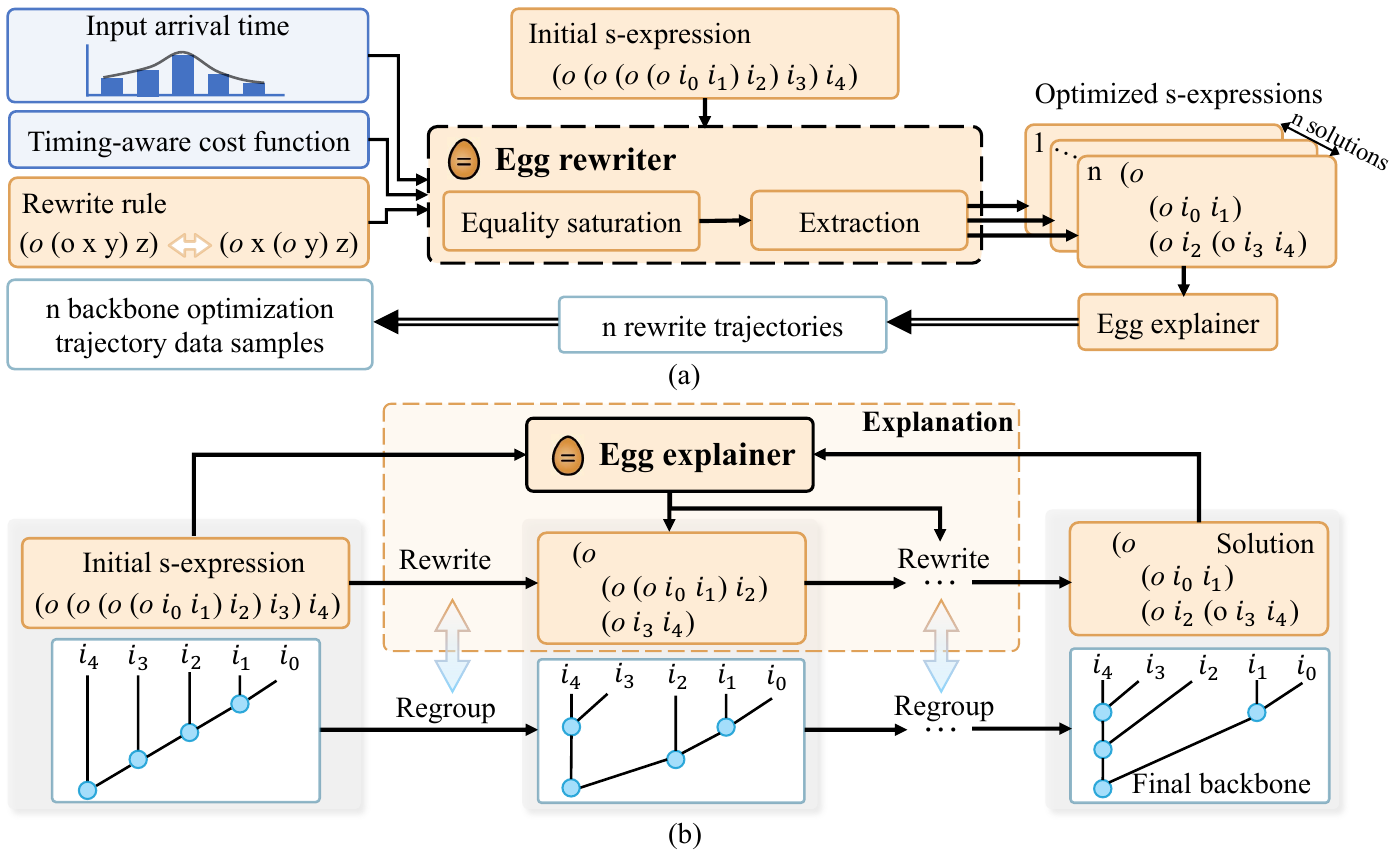}

\caption{E-graph based optimization trajectory dataset generation: (a) Given diverse input arrival profiles and extraction strategies, \texttt{egg} extracts a variety of optimized s-expressions. (b) Each optimized s-expression is explained as a rewrite trajectory using \texttt{egg} explainer, then forming interpretable regroup sequences for backbone optimization.}

\label{fig:method:datagen}
\end{figure*}

Optimizing prefix adder backbones under specific delay constraints and non-uniform input arrival times remains highly complex.
For LRMs, this task constitutes an unseen scenario, highlighting the necessity of high-quality optimization trajectory data for effective fine-tuning.
To enable effective fine-tuning of the LRM, we propose an e-graph-based data generation framework, as illustrated in \Cref{fig:method:datagen} and \Cref{fig:method:datagen2}.
This data generation flow consists of four main components:

(1) utilizing e-graphs to derive diverse backbone solutions tailored to distinct input arrival profiles and bitwidth configurations;

(2) leveraging the explanation provided by \texttt{egg} to produce interpretable optimization rewrite (i.e., regroup function callings in \textit{Prefix Agent} Phase~ I) trajectories derived from the backbone solutions;

(3) employing commercial LRMs such as GPT-o3~\cite{openai2025o3} to generate structured \texttt{<think>...</think>} reasoning content based on the rewrite trajectories; and

(4) integrating the \textit{PrefixAgent} prompts with tool feedback, as well as the regroup function callings trajectories, to generate comprehensive training samples for the fine-tuning of LRM.

\subsection{Backbone Representation in E-graphs}

The \texttt{egg} library represents expressions using nested s-expressions format.
An s-expression (short for symbolic expression) is a parenthesis-based notation commonly used to represent nested hierarchical structures, originally popularized in Lisp programming.
In the context of e-graphs, s-expressions describe computation trees in a fully parenthesized form, where both operators and operands must be explicitly defined in a domain-specific intermediate language.
To describe prefix adder backbones, we define a domain-specific intermediate language, \texttt{BackboneLang}, to encode backbone structures in a structured and concise form.
In \texttt{BackboneLang}, each input node is represented as a \texttt{Leaf} annotated with its input arrival time.
The language supports only one single binary operator \texttt{\(\mathit{o}\)}, which denotes a grouping operation between two operands—either leaf nodes or previously grouped sub-expressions.
For example, grouping inputs \(i_0\) and \(i_1\) is expressed as \texttt{(\(\mathit{o}\) i0 i1)}, where \(i_0\) and \(i_1\) are \texttt{Leaf} nodes.
In this way, the backbone structure corresponding to \Cref{eq:backbone1} can be represented in \texttt{BackboneLang} as:

\begin{dslbox}{BackboneLang Example1}
(o
  (o
    (o 
      (o i0 i1) 
      (o i2 i3))
    (o i4 i5) 
  (o i6 i7)))
\end{dslbox}

\subsection{Backbone Rewrites in E-graphs}

We define a single associativity rewrite rule as follows:
\begin{equation}
(\mathit{o}\;(\mathit{o}\;\texttt{x}\;\texttt{y})\;\texttt{z})\;\Longleftrightarrow\;(\mathit{o}\;\texttt{x}\;(\mathit{o}\;\texttt{y}\;\texttt{z}))
\tag{R1}\label{eq:assoc}
\end{equation}
This associativity rewrite rule on S-expressions is semantically equivalent to a regroup operation in the prefix backbone.
That is, rewriting an expression according to rule \ref{eq:assoc} corresponds exactly to restructuring the backbone by modifying the regroup of two subtrees.
For example, in \Cref{fig:method:regroup}, the rewrite operation transforms the expression from \Cref{eq:backbone1} to \Cref{eq:backbone2}, resulting in an alteration of the corresponding s-expression to:

\begin{dslbox}{BackboneLang Example2}
(o
  (o 
    (o i0 i1) 
    (o i2 i3))
  (o 
    (o i4 i5) 
    (o i6 i7)))
\end{dslbox}

Starting from an initial fully serial backbone expression, we employ \texttt{egg}'s equality saturation~\cite{Egraph-2009POPL-Tate} to apply the associativity rewrite rule~\ref{eq:assoc}.
This process compactly encodes the complete solution space of backbone structures within a single e-graph \(\mathcal{G}\).

\subsection{Timing-aware Cost Function for Extraction}
\label{subsec:cost_func}

From the generated e-graph \(\mathcal{G}\), we extract multiple backbone candidates under given input arrival times and delay constraints.
The bottom-up extractor in \texttt{egg} assumes a compositional cost model, where the cost of each enode is computed from its operator and the costs of its children.
The built-in costs in \texttt{egg} (AST size or AST depth) are not well aligned with our backbone setting: for a fixed bit-width, the backbone node count is constant, so an AST-size cost becomes degenerate, and a pure depth cost cannot reflect bit-wise non-uniform input arrival times.
We therefore employ the \texttt{egg} extractor with a timing-aware cost tailored to the backbone structure.

Prior work \cite{DSE-2019TCAD-MA,Datapath-2024ICCAD-Zuo} has shown that a linear approximation provides estimates of path delays in prefix adders with high fidelity.
Given that each backbone node has a fixed fanout count of one, the delay is primarily determined by the longest path in the binary tree structure.
Consequently, we define the delay cost as a linear function of the path length, $y = wx + b$, where $x$ denotes the number of nodes on the path.
Accordingly, we define the cost function $C$ to guide e-graph extraction under non-uniform input arrival times.
This function recursively computes the cost of each internal node based on the arrival times of its inputs and the linear delay model:

\begin{equation}
C(n) =
\begin{cases}
t_i, & \text{if } n = \texttt{Leaf}(x_i); \\[4pt]
\max(C(l),\ C(r)) + d + \lambda, & \text{if } n = \texttt{Op}(l, r).
\end{cases}
\label{eq:arrival-time-cost}
\end{equation}
Here, \(t_i\) denotes the input arrival time at \texttt{Leaf} \(x_i\), \(d\) is a constant node delay, and \(\lambda\) is a timing margin used to model physical-implementation uncertainty.

To evaluate the impact of the cost design, we compare our timing-aware cost against two baselines on a 16-bit backbone.
For two representative non-uniform arrival profiles, we extract 100 solutions and record the best critical-path delay under three cost functions: the proposed timing-aware cost, an AST-depth cost, and a random baseline.
As summarized in Table~\ref{table:cost_sensitivity}, the timing-aware cost consistently yields backbones with smaller critical-path delays than both baselines, which indicates that it is a suitable and effective extractor-compatible model in our setting.

\begin{table}[!tb]
\caption{Best critical-path delay (ns) under different extraction cost functions}
\centering

\begin{tabular}{|c|c|c|c|}
\hline
 & Timing-aware& AST-depth & Random \\
\hline
profile~1 & 0.4862 & 0.5684 & 0.6692 \\
profile~2 & 0.4680 & 0.5514 & 0.6851 \\
\hline
\end{tabular}
\label{table:cost_sensitivity}

\end{table}

\begin{figure*}[!t]
  \centering
\includegraphics[width=1.\linewidth]{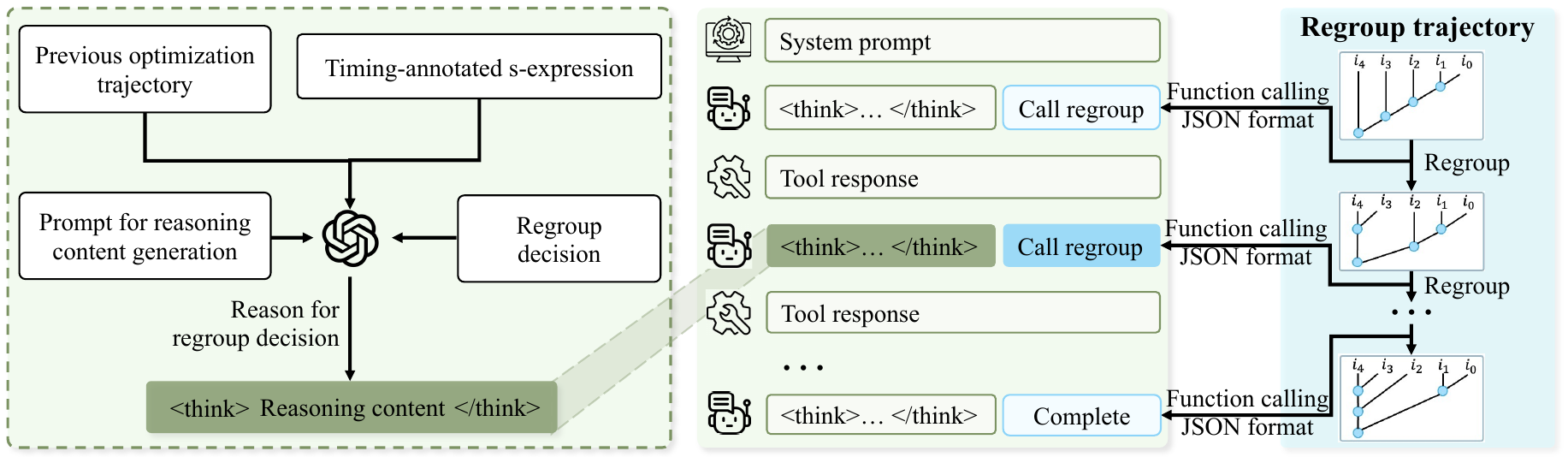}
\caption{Fine tuning data synthesis}
\label{fig:method:datagen2}
\end{figure*}

\subsection{Optimization Trajectory Generation}
\label{subsec:opt_traj}
The cost function satisfies decomposability and monotonicity, two essential properties required by \texttt{egg}'s optimal extraction algorithm.
We therefore adopt the default bottom-up extractor provided by \texttt{egg} to retrieve the optimal backbone solution.
The optimal solution corresponds to the tightest timing constraint scenario.
To emulate more general area–delay trade-off conditions, we additionally generate suboptimal solutions by introducing controlled perturbations during the extraction process.
This modification allows the extractor to explore a broader set of feasible backbone structures that reflect a wider range of relaxed timing constraints.

As illustrated in \Cref{fig:method:datagen}~(a), we first generate numerous backbone solutions (i.e., optimized s-expressions) using \texttt{egg}'s extraction mechanism. Subsequently, for each of these solutions, the \texttt{egg} explainer is utilized to produce an explanation trajectory in the form of a rewrite sequence, describing how the backbone structure is transformed from the initial serial expression.
As illustrated in \Cref{fig:method:datagen}~(b), each rewrite of the s-expression within a trajectory directly corresponds to a \texttt{regroup} operation on the backbone.
Consequently, these sequences serve as interpretable optimization trajectories, explicitly capturing the chain of \texttt{regroup} operations applied to reach a given backbone structure.
Among all backbone solutions encoded in the e-graph, we select training samples that yield minimal increases in the final prefix adder level \(L\) relative to the backbone level \(L_B\).
Such samples are more likely to correspond to zero-deficiency or low-deficiency adders.
A summary of the collected trajectory statistics across bit-widths is given in \Cref{table:trajectory_stats}.
The arrival profiles span seven families (linear, convex, concave, MSB-first, LSB-first, trapezoidal, and random), with multiple instances per family. Equality saturation encodes the full single-Catalan grouping space for each bit-width, and timing-aware extraction samples a structurally diverse subset, as reflected by the backbone-level ranges in \Cref{table:trajectory_stats}. We then perform an approximately 8:1:1 split, stratified by profile family, at the level of individual profile instances, so that a profile instance used for evaluation never appears in training, even when other instances of the same family do.

\begin{table}[!tb]
\caption{Trajectory and profile statistics (\#Profile = input arrival profiles, \#Traj. = trajectories, Length = trajectory length, Level = backbone level, Target delay = target delay range, Steps = total regroup steps).}
\centering
\begin{tabular}{|c|c|c|c|c|c|c|}
\hline
 & \#Profile & \#Traj. & Length & Level & Target delay (ns) & Steps \\
\hline
 8-bit  & 23 & 135 & 3--6   & 3--6   & 0.4--0.7 & 608  \\
12-bit  & 21 & 132 & 4--8   & 4--7   & 0.5--0.8 & 818  \\
16-bit  & 20 & 128 & 12--18 & 4--9   & 0.5--0.8 & 1920 \\
24-bit  & 18 & 120 & 17--26 & 5--12  & 0.6--1.0 & 2580 \\
32-bit  & 16 & 110 & 24--37 & 5--15  & 0.7--1.2 & 3315 \\
48-bit  & 16 & 100 & 41--61 & 6--18  & 0.8--1.4 & 4984 \\
64-bit  & 16 &  90 & 49--84 & 6--25  & 1.0--1.7 & 5760 \\
\hline
\end{tabular}
\label{table:trajectory_stats}
\end{table}

\subsection{Data Synthesis and Fine Tuning}
With the optimization trajectories established, synthesizing complete fine-tuning samples for the LRM necessitates augmenting each \texttt{regroup} decision in the trajectory with explicit Chain-of-Thought (CoT) reasoning, which is annotated within \texttt{<think>...</think>} tags.
As illustrated on the left side of \Cref{fig:method:datagen2}, we synthesize the CoT reasoning content for each \texttt{regroup} decision.
This is achieved by prompting a powerful teacher LRM, such as GPT-o3~\cite{openai2025o3}, with the historical optimization trajectory and the specific \texttt{regroup} choice for the current step, instructing it to generate a plausible reasoning chain.
To assemble the final fine-tuning data for an agent's response, the \texttt{regroup} decision is formatted as a JSON-style function call.
This call is then concatenated with the corresponding synthesized reasoning content, which is encapsulated within \texttt{<think>...</think>} tags.
Example data sample for fine-tuning is shown as below:

\begin{promptbox}{Backbone Optimization Trajectory Data Example}
System prompt: \\
You are an agent for backbone optimization ..... + [input arrival time profile] + [target delay constraint] + [initial timing-annotated s-expression] + [instructions]

PrefixAgent: \\
<think>Now we need to optimize the adder backbone to 2ns ... First we look at the input arrival profile ... It is the most critical part, we need to add parallelism at ... So we need to regroup node (2,2) and (1,0) </think>regroup tool calls \\

Tool response: \\
Regroup success; Regroup success + [logic level improvement] + [delay improvement] + [current timing-annotated s-expression]\\

<think>... We still need to speed up at ... </think>regroup tool calls

Tool response: Regroup success + [logic level improvement] + [delay improvement] + [current timing-annotated s-expression]\\

<think> ... Based on this analysis, I decide to regroup (15,14) and (13,10)</think>regroup tool calls \\

....

Tool response: Regroup success + [logic level improvement] + [delay improvement] + [current s-expression]\\

<think>The current critical path delay already satisfies the timing constraint ... So we can complete now.</think>Call complete


\end{promptbox}

\begin{figure*}[!t]
    \centering
    \definecolor{YellowBorder}{RGB}{223,161,89}
\definecolor{BlueBorder}{RGB}{30,95,225}
\definecolor{PinkBorder}{RGB}{235, 70, 90}
\definecolor{GreenBorder}{RGB}{91,150,60}
\definecolor{PrefixGPTgray}{RGB}{68,68,68}

\begin{tikzpicture}
\begin{groupplot}[group style={group size= 3 by 1, horizontal sep=0.86cm, group name=myplot}, height=4.68cm, width=5cm]
\nextgroupplot[minor tick num=0,
xlabel={Area ($\upmu$m$^2$)},
ylabel={Delay (ns)},
y label style={at={(-0.2,0.5)}},
ylabel near ticks,
legend style={
    draw=none,
	at={(0.25,1.)},
	anchor=north,
	legend columns=-1,
}
]
    \addplot[style={mark=star, mark size=1.4pt, line width=1.2pt, draw=BlueBorder}]  table [x=area, y=delay, col sep=comma] {pgfplot/data/processed/mcts16.csv};
    \addplot[style={mark=o, mark size=1.4pt, line width=1.2pt, draw=PinkBorder}]  table [x=area, y=delay, col sep=comma] {pgfplot/data/processed/rl16.csv};
    \addplot[style={mark=triangle, mark size=1.4pt, line width=1.2pt, draw=GreenBorder}]  table [x=area, y=delay, col sep=comma] {pgfplot/data/processed/vae16.csv};
    \addplot[style={mark=x, mark size=1.6pt, line width=1.2pt, draw=PrefixGPTgray}]  table [x=area, y=delay, col sep=comma] {pgfplot/data/processed/prefixgpt16.csv};
    \addplot[style={mark=*, mark size=1.4pt, line width=1.2pt, draw=YellowBorder}]  table [x=area, y=delay, col sep=comma] {pgfplot/data/processed/ours16.csv};
\coordinate (left) at (rel axis cs:0,1);

\nextgroupplot[minor tick num=0,
xlabel={Area ($\upmu$m$^2$)},
y label style={at={(-0.2,0.5)}},
xtick={400,460,520},
ylabel near ticks,
legend style={
    draw=none,
	at={(0.5,1.22)},
	anchor=north,
	legend columns=-1,
}
]
    \addplot[style={mark=star, mark size=1.4pt, line width=1.2pt, draw=BlueBorder}]  table [x=area, y=delay, col sep=comma]{pgfplot/data/processed/mcts32.csv};\addlegendentry{MCTS\cite{Datapath-2024NIPS-Lai}};
    \addplot[style={mark=o, mark size=1.4pt, line width=1.2pt, draw=PinkBorder}]  table [x=area, y=delay, col sep=comma] {pgfplot/data/processed/rl32.csv};\addlegendentry{PrefixRL\cite{RL-2021DAC-Roy}};
    \addplot[style={mark=triangle, mark size=1.4pt, line width=1.2pt, draw=GreenBorder}]  table [x=area, y=delay, col sep=comma] {pgfplot/data/processed/vae32.csv};\addlegendentry{CircuitVAE\cite{Datapath-2024DAC-Song}};
    \addplot[style={mark=x, mark size=1.6pt, line width=1.2pt, draw=PrefixGPTgray}]  table [x=area, y=delay, col sep=comma] {pgfplot/data/processed/prefixgpt32.csv};\addlegendentry{PrefixGPT\cite{Datapath-2026AAAI-Ding}};
    \addplot[style={mark=*, mark size=1.4pt, line width=1.2pt, draw=YellowBorder}]  table [x=area, y=delay, col sep=comma] {pgfplot/data/processed/ours32.csv};\addlegendentry{PrefixAgent};
\coordinate (mid) at (rel axis cs:0.5,1);

\nextgroupplot[minor tick num=0,
xlabel={Area ($\upmu$m$^2$)},
y label style={at={(-0.2,0.5)}},
ylabel near ticks,
legend style={
    draw=none,
	at={(0.9,1.4)},
	anchor=north,
	legend columns=-1,
}
]
    \addplot[style={mark=star, mark size=1.4pt, line width=1.2pt, draw=BlueBorder}]  table [x=area, y=delay, col sep=comma] {pgfplot/data/processed/mcts64.csv};
    \addplot[style={mark=o, mark size=1.4pt, line width=1.2pt, draw=PinkBorder}]  table [x=area, y=delay, col sep=comma] {pgfplot/data/processed/rl64.csv};
    \addplot[style={mark=triangle, mark size=1.4pt, line width=1.2pt, draw=GreenBorder}]  table [x=area, y=delay, col sep=comma] {pgfplot/data/processed/vae64.csv};
    \addplot[style={mark=x, mark size=1.6pt, line width=1.2pt, draw=PrefixGPTgray}]  table [x=area, y=delay, col sep=comma] {pgfplot/data/processed/prefixgpt64.csv};
    \addplot[style={mark=*, mark size=1pt, line width=1.2pt, draw=YellowBorder}]  table [x=area, y=delay, col sep=comma] {pgfplot/data/processed/ours64.csv};
\coordinate (right) at (rel axis cs:1,1);

\end{groupplot}
\path (left)--(right) coordinate[midway] (group center);
\path (myplot c2r1.north west|-current bounding box.north)--
coordinate(legendpos)
(myplot c2r1.north west|-current bounding box.north);
\end{tikzpicture}
    \caption{Pareto-frontiers under uniform arrival time. From left to right: 16-bit; 32-bit; 64-bit.}
    \label{fig:results}
\end{figure*}

Highly repetitive CoT templates can limit the model’s ability to generalize.
In such cases, the model tends to overfit to surface-level textual profiles rather than attend to structural features.
This behavior results in reduced sensitivity to function-call accuracy, despite achieving low training loss that does not reflect actual reasoning fidelity.
To jointly supervise reasoning generation and structural function calling decision accuracy, we define a composite training objective:
\[
L_{\mathrm{total}} = L_{\mathrm{CoT}} + \lambda \cdot L_{\mathrm{function\_call}}
\]
Here, \(L_{\mathrm{CoT}}\) supervises the generation of coherent chain-of-thought reasoning, while \(L_{\mathrm{function\_call}}\) promotes accurate invocation of regroup via function calls.

\section{Experimental Results}

\label{sec:exp}
\subsection{Setup}
\label{subsec:setup}

\begin{table*}[!tb]
\caption{Comparison of non-uniform arrival time profiles}
\centering
\resizebox{\columnwidth}{!}{%
\begin{tabular}{|c|c|cc|cc|cc|cc|}
\hline
\multirow{2}{*}{Input arrival profile} & \multirow{2}{*}{Method} & \multicolumn{2}{c|}{16-bit} & \multicolumn{2}{c|}{32-bit} & \multicolumn{2}{c|}{54-bit} & \multicolumn{2}{c|}{64-bit} \\
\cline{3-10}
& & Area ($\upmu$m$^2$) & Delay (ns) & Area ($\upmu$m$^2$) & Delay (ns) & Area ($\upmu$m$^2$) & Delay (ns) & Area ($\upmu$m$^2$) & Delay (ns) \\
\hline
\multirow{6}{*}{LSB-first}
& DP\cite{Datapath-2007GLSVLSI-Matsunaga} & 236 & 0.6089 & 461 & 0.8109 & 761 & 0.9540 & 1130 & 1.5307 \\
& MCTS\cite{Datapath-2024NIPS-Lai}        & 217 & 0.5810 & 438 & 0.8012 & 707 & 0.9531 & 1045 & 1.5123 \\
& PrefixRL\cite{RL-2021DAC-Roy}           & 222 & 0.5928 & 445 & 0.8035 & 726 & 0.9486 & 1078 & 1.4984 \\
& PrefixGPT\cite{Datapath-2026AAAI-Ding} & \textbf{215} & 0.6014 & 438 & 0.8012 & 712 & 0.9489 & 1051 & 1.5105 \\
& CircuitVAE\cite{Datapath-2024DAC-Song}  & 219 & 0.5948 & 442 & 0.7946 & 735 & 0.9584 & 1074 & 1.5078 \\
& \textbf{PrefixAgent}                    & 217 & \textbf{0.5810} & \textbf{426} & \textbf{0.7980} & \textbf{697} & \textbf{0.9461} & \textbf{938} & \textbf{1.4882} \\
\hline
\multirow{6}{*}{Random}
& DP\cite{Datapath-2007GLSVLSI-Matsunaga} & 220 & 0.5740 & 490 & 0.8607 & 739 & 1.1249 & 1134 & 1.5610 \\
& MCTS\cite{Datapath-2024NIPS-Lai}        & 207 & 0.5567 & 463 & 0.8509 & 668 & 0.9963 & 1075 & 1.5591 \\
& PrefixRL\cite{RL-2021DAC-Roy}           & 212 & 0.5543 & 463 & 0.8509 & 679 & 1.0986 & 1104 & 1.5643 \\
& PrefixGPT\cite{Datapath-2026AAAI-Ding} & 207 & 0.5567 & 449 & 0.8560 & 664 & 1.0587 & 1094 & 1.4915 \\
& CircuitVAE\cite{Datapath-2024DAC-Song}  & 210 & 0.5478 & 472 & 0.8489 & 672 & 0.9960 & 1097 & 1.5630 \\
& \textbf{PrefixAgent}                    & \textbf{205} & \textbf{0.5450} & \textbf{440} & \textbf{0.8373} & \textbf{653} & \textbf{0.9980} & \textbf{953} & \textbf{1.5579} \\
\hline
\end{tabular}
\label{table:nu}
}
\end{table*}

Both fine-tuning and inference are conducted on a single node with 8 NVIDIA H100 (80 GB) GPUs and 2 TB of host memory.
The \textit{PrefixAgent} framework adopts the QwQ-32B~\cite{alibabaQwQ32B} model as the LRM, which exhibits strong reasoning and function-calling capabilities. We fine-tune it using LoRA with a per-GPU batch size of 2 for three epochs, reaching a peak GPU memory of about 60 GB and completing in roughly 12 hours.
We collect CoT content using the GPT-o3~\cite{openai2025o3} model (API snapshot \texttt{o3} in a high-entropy, tool-free configuration:
\texttt{temperature}~$=0.8$, \texttt{top\_p}~$=0.95$, \texttt{max\_output\_tokens}~$=16384$, \texttt{seed}~$=7$, \texttt{tools}~$=[]$, and \texttt{tool\_choice}~$=\text{"none"}$).
We use NanGate45 Open Cell Library~\cite{nangate45} and perform logic synthesis with Yosys~\cite{Yosys}, and physical implementation with OpenROAD~\cite{OpenROAD-2019DAC-Ajayi}. Functional correctness of all generated adders is verified by equivalence checking using ABC~\cite{Berkeley-ABC}.
Our baselines include:

\textbf{DP}~\cite{Datapath-2007GLSVLSI-Matsunaga}: A dynamic-programming method that minimizes area under non-uniform arrival times.

\textbf{PrefixRL}~\cite{RL-2021DAC-Roy}: A RL-based optimization approach, reimplemented following the methodology described in~\cite{RL-2021DAC-Roy}.

\textbf{MCTS}~\cite{Datapath-2024NIPS-Lai}: A MCTS-based method, using the open-source codes from~\cite{Datapath-2024NIPS-Lai}.

\textbf{CircuitVAE}~\cite{Datapath-2024DAC-Song}: A generative approach based on VAE. We re-implement the model following~\cite{Datapath-2024DAC-Song}, incorporating the optimization strategy proposed in~\cite{10.5555/3495724.3496669}. Under each configuration, randomly sample 1000 initial adders from PrefixRL solutions as initial structures.

\textbf{PrefixGPT}~\cite{Datapath-2026AAAI-Ding}: A transformer-based prefix-adder generation method, extended to 64-bit in our flow.

We also attempted to reproduce PrefixLLM~\cite{Datapath-2024Arxiv-Xiao}; however, it consistently generated ripple-carry adders or replicated the given initial structures, and failed to produce valid new prefix structures for cases beyond 16 bits.
Therefore, we exclude it from comparative evaluation.
Detailed discussion of PrefixLLM can be found in \Cref{subsec:discussion}.

We evaluate all methods under both uniform and non-uniform arrival time profile conditions.
We set various target delays for area-delay trade-off scenarios and conduct physical implementation with the target delays.
All approaches are evaluated under the same delay target to ensure a fair comparison.
For RL, MCTS, and VAE baselines, we adopt a unified cost formulation to enable consistent timing–area trade-off evaluation.
Specifically, the cost formulation penalizes timing violations and rewards area reduction.
We further evaluate \textit{PrefixAgent} under a commercial physical design flow.

\subsection{Comparing with Baselines}
\label{subsec:results}

\begin{table*}[!tb]
\caption{Ablation study on framework components, prefix-graph encoding, and training-data scaling. The full PrefixAgent (bottom row) uses the EPR encoding and the complete training set. All runs use the LSB-first profile under a common delay target.}
\centering
\resizebox{\columnwidth}{!}{%
\begin{tabular}{|c|cc|cc|cc|}
\hline
\multirow{2}{*}{Method} & \multicolumn{2}{c|}{16-bit} & \multicolumn{2}{c|}{32-bit} & \multicolumn{2}{c|}{64-bit} \\
\cline{2-7}
& Area ($\upmu$m$^2$) & Delay (ns) & Area ($\upmu$m$^2$) & Delay (ns) & Area ($\upmu$m$^2$) & Delay (ns) \\
\hline
Phase~I only                         & 198 & 0.6512 & 407 & 0.8363 &  890 & 1.5215 \\
Phase~II w/ grid encoding           & 223 & 0.5987 & 445 & 0.8106 & 1046 & 1.5300  \\
Phase~II w/ adjacency-list encoding & 227 & 0.6045 & 457 & 0.7940 & 1067 & 1.5482 \\
Phase~II w/o critical-path prompts   & 217 & 0.5810 & 439 & 0.7683 &  960 & 1.4958 \\
w/o CoT reasoning                    & 245 & 0.6792 & 516 & 0.8245 & 1176 & 1.5520 \\
\hline
PrefixAgent w/o fine-tuning & 225 & 0.5867 & 464 & 0.8106 & 1106 & 1.5015 \\
PrefixAgent (1/4 data)      & 221 & 0.5914 & 448 & 0.8134 & 1064 & 1.5150 \\
PrefixAgent (1/2 data)      & 217 & 0.5810 & 437 & 0.7945 & 951 & 1.5069 \\
\hline
Phase~I + Phase~II (PrefixAgent)     & 217 & 0.5810 & 426 & 0.7980 & 938 & 1.4882 \\
\hline
\end{tabular}
\label{table:ablation1}
}
\end{table*}

\begin{figure}[!tb]
\centering
\includegraphics[width=0.7\linewidth]{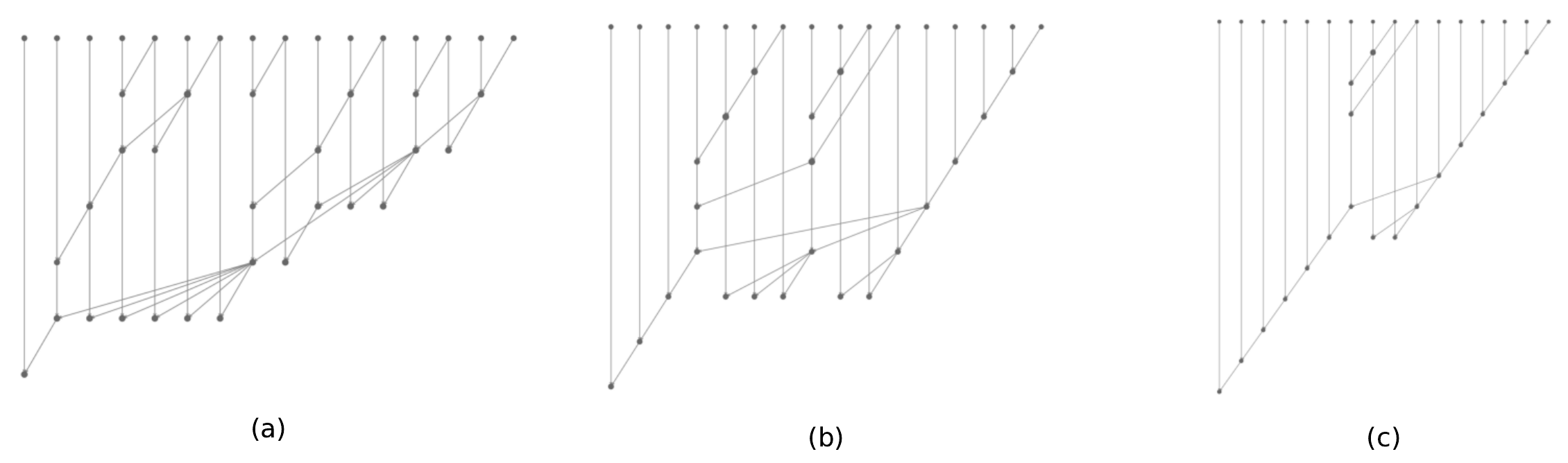}
\caption{Representative 16-bit prefix structures along the Pareto frontier: (a) delay-optimized; (b) balanced; (c) area-optimized.}
\label{fig:exp:example2}
\end{figure}

Under the uniform arrival time setting, we evaluate each method across six target delay values per bit-width, representing a range of area–delay trade-off scenarios.
For each target delay, \textit{PrefixAgent} generates a single optimized adder structure.
In contrast, each baseline (MCTS, RL, VAE, and PrefixGPT) is executed with 5000 samples per delay target, yielding 30,000 designs per method.
The resulting Pareto fronts are illustrated in \Cref{fig:results}. Across all tested bit-widths, \textit{PrefixAgent} consistently achieves high-quality solutions.
Notably, in the 64-bit case, it Pareto-dominates all baseline approaches.
To show how these trade-offs are realized at the structural level, \Cref{fig:exp:example2} presents three representative 16-bit designs sampled along the Pareto frontier. The delay-optimized design adopts more prefix nodes and a shallower, more parallel topology to reduce the logic level. The area-optimized design uses fewer nodes and a more serial structure to save area, and the balanced design lies between them. This illustrates how \textit{PrefixAgent} traverses the area–delay space through interpretable changes in prefix topology.

\begin{figure}[!tb]
\centering
\includegraphics[width=0.5\linewidth]{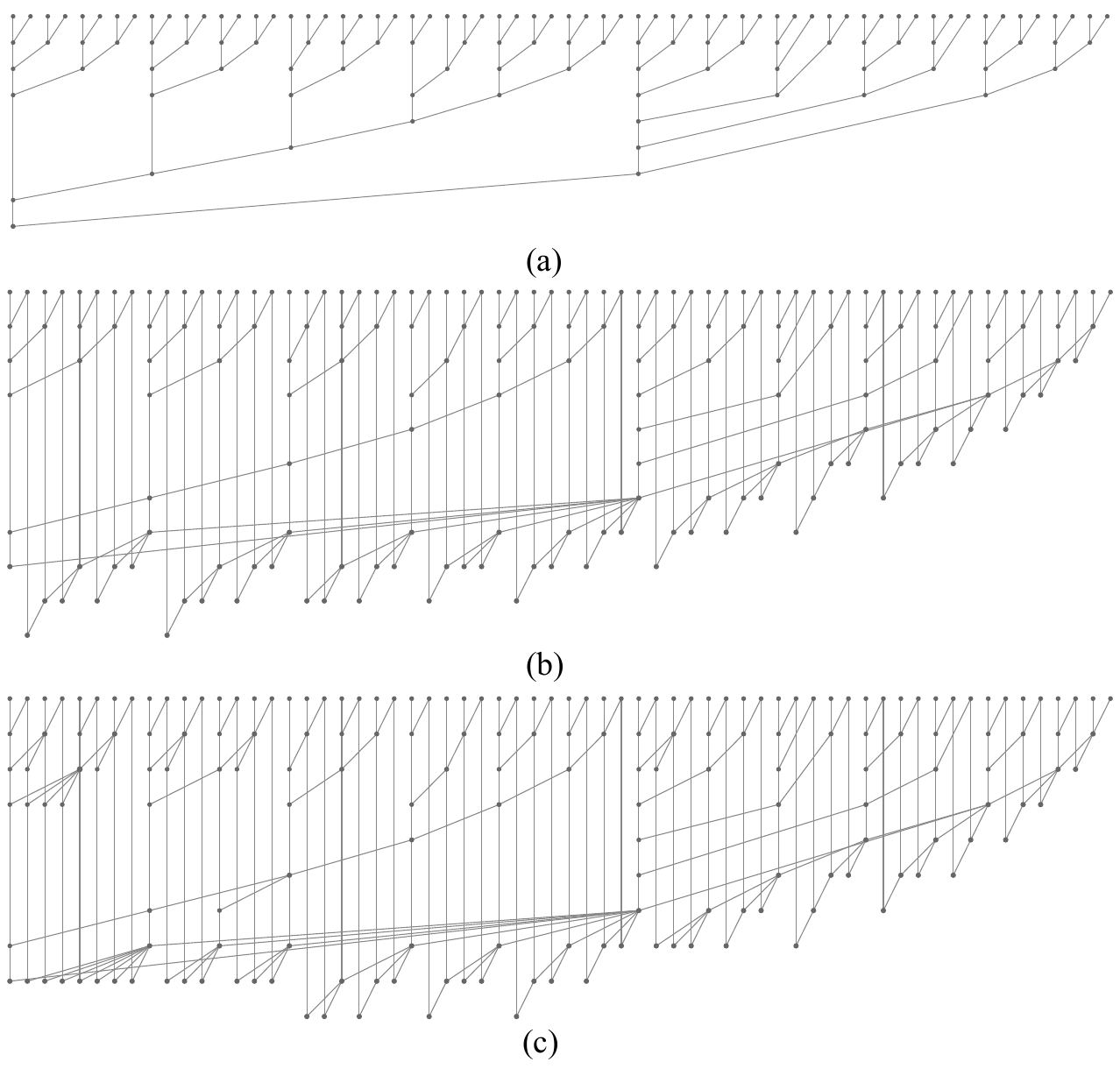}
\caption{Example structures generated by \textit{PrefixAgent}: (a) optimized backbone; (b) completed adder after Phase~ I; (c) final optimized adder after Phase~ II.}
\label{fig:exp:example}
\end{figure}

For the non-uniform arrival time scenario, we evaluate two distinct arrival time profiles: (i) \textit{LSB-first}, where the least significant half of the input bits arrive earlier, and (ii) \textit{Random}, where arrival times are randomly assigned across all bits.
For each arrival profile and bit-width configuration, we set a fixed target delay.
The specific LSB-first and Random profiles used here are taken from the held-out test partition (\Cref{subsec:opt_traj}); no instance of either appears in the training set.
Each learning-based baseline samples 5000 candidate designs under the given delay constraint, while the dynamic-programming baseline is deterministic and yields a single solution per configuration.
Among them, we select the adders with the smallest area that meet the timing constraints, allowing at most 1\% timing violation.
The comparison results are summarized in \Cref{table:nu}.\footnote{At smaller bit-widths, different methods can converge to the same prefix structure and therefore report identical post-synthesis area and delay.}
As shown, \textit{PrefixAgent} achieves the lowest area in seven of the eight configurations; the only exception is the 16-bit LSB-first case, where PrefixGPT attains a marginally smaller area.
Notably, under the 64-bit configuration, it achieves up to 11.3\% area reduction compared to the best-performing baseline.
These results indicate that PrefixGPT remains competitive at smaller bit-widths, while the advantage of \textit{PrefixAgent} becomes more pronounced at larger bit-widths and across the broader set of evaluated scenarios.
In addition, the 54-bit configuration does not appear in the fine-tuning dataset (\Cref{table:trajectory_stats}); \textit{PrefixAgent} nonetheless attains the smallest area under both profiles, providing evidence of transfer to this unseen bit-width.
Since 54 bits is the only unseen bit-width evaluated and both arrival-profile families appear in the training distribution, these results do not establish broader generalization to arbitrary bit-widths or unseen profile families.
Example structures generated by \textit{PrefixAgent} are shown in \Cref{fig:exp:example}.

We report the average runtime of PrefixAgent and the MCTS, RL, and VAE baselines in \Cref{fig:runtime}. Compared to baseline approaches that rely on extensive search within a large design space, \textit{PrefixAgent} performs targeted optimizations guided by LRM decisions.
As a result, it achieves over 10$\times$ speedup across all evaluated configurations.
The significant runtime advantage of \textit{PrefixAgent} stems from its ability to generalize across design cases without retraining.
Baseline methods such as RL, MCTS, and VAE require retraining for each new configuration, including variations in bit-width, input arrival time profiles, and target delay constraints.
In contrast, \textit{PrefixAgent} is trained once on a carefully curated dataset covering a broad distribution of design scenarios.
During inference, the LRM leverages its reasoning capabilities and the structured decomposition of the complex task into a backbone optimization task.
This one-time training paradigm, combined with problem space reduction via backbone abstraction, substantially reduces the per-case optimization cost and enhances scalability.

\begin{figure}[!tb]
    \centering
    \definecolor{YellowBorder}{RGB}{223,161,89}
\definecolor{YellowFill}{RGB}{254,236,214}
\definecolor{DeepGreenBorder}{RGB}{192,201,178}
\definecolor{LightGreenFill}{RGB}{243,245,240}
\definecolor{BlueBorder}{RGB}{69,96,160}
\definecolor{PinkText}{RGB}{164,42,37}
\definecolor{PinkBorder}{RGB}{235, 70, 90}
\definecolor{GreenBorder}{RGB}{91,130,60}

\pgfplotsset{
	width =0.516\linewidth,
	height=0.326\linewidth
}

\begin{tikzpicture}
\begin{axis}[
    ybar,
    xticklabels={16-bit, 32-bit, 64-bit},xtick={1.0,3.0,5.0},
    xtick align=inside,
    ylabel={Times (s)},
    ylabel near ticks,
    bar width = 8pt,
    xmin=0,
    xmax=5.6,
    ymin=0,
    ymax=72000,
    legend style={at={(0.5,1.32)},
    draw=none,anchor=north,legend columns=-1}
    ]

\addplot +[ybar, fill=BlueBorder!85!white,   draw=BlueBorder!85!white, area legend] table [x={x},  y={mcts}] {hv.dat};
\addplot +[ybar, fill=PinkBorder!85!white,   draw=PinkBorder!85!white, area legend] table [x={x},  y={rl}] {hv.dat};
\addplot +[ybar, fill=GreenBorder!85!white,,   draw=GreenBorder!85!white, area legend] table [x={x},  y={vae}] {hv.dat};
\addplot +[ybar, fill=YellowBorder!94!white,   draw=YellowBorder!94!white, area legend] table [x={x},  y={ours}] {hv.dat};

\legend{MCTS\cite{Datapath-2024NIPS-Lai}, RL\cite{RL-2021DAC-Roy}, VAE\cite{Datapath-2024DAC-Song}, PrefixAgent}
\end{axis}
\end{tikzpicture}
    \caption{Average runtime comparison.}
    \label{fig:runtime}
\end{figure}

\begin{table}[!t]
\centering

\caption{Prefix-graph reasoning accuracy (\%) under different representations. Each probe consists of 100 questions over randomly sampled prefix graphs.}
\begin{tabular}{|l|c|c|c|}
\hline
Probe & EPR & Grid & Adjacency-list \\
\hline
fanout\_cone\_size & 94 & 59 & 36 \\
reachable          & 96 & 63 & 47 \\
fanin\_cone\_size  & 97 & 58 & 33 \\
\hline
\end{tabular}
\label{table:repr_reasoning}

\end{table}

\begin{table}[!tb]
\caption{Ablation study on CoT reasoning}
\centering
\begin{tabular}{|c|c|c|c|c|}
\hline
 & Model & Area ($\upmu$m$^2$) & Delay (ns) & \#Tool Calls \\
\hline
\multirow{2}{*}{16-bit}
 & GPT-o3 & 217 & 0.5810  & 31 \\
 & Qwen3 & 217 & 0.5976 &  30\\
\hline
\multirow{2}{*}{32-bit}
 & GPT-o3 & 426  & 0.7980  & 51 \\
 & Qwen3 &  426  &  0.7980 &  51\\
\hline
\multirow{2}{*}{64-bit}
 & GPT-o3 & 938 & 1.4882 & 73 \\
 & Qwen3 & 954 & 1.4957 &  76\\
\hline
\end{tabular}

\label{table:cot_robust}

\end{table}

\subsection{Ablation Studies}
\label{subsec:ablation}

We conduct five ablation studies: (i) the contribution of framework components, namely Phase~II refinement and the critical-path prompts; (ii) the Phase~II prefix-graph encoding, comparing EPR against grid and adjacency-list encodings, for which we additionally report a reasoning-probe analysis; (iii) the effect of CoT reasoning against pure tool-call supervision; (iv) the effect of training-data scale, from no fine-tuning to the full training set; and (v) robustness to the choice of LRM for CoT collection. Items (i)--(iv) are reported in \Cref{table:ablation1}, the supporting reasoning-probe analysis in \Cref{table:repr_reasoning}, and item (v) in \Cref{table:cot_robust}. All ablations use the LSB-first arrival profile. Except for the Phase~I-only and w/o CoT reasoning settings, the realized delays in \Cref{table:ablation1} are close, so the remaining comparisons are read on area.

As shown in \Cref{table:ablation1}, in Phase~I, the complete adders from the backbone may violate timing due to increased logic depth or fanout introduced
by the auxiliary nodes. Phase~ II performs timing-oriented refinements and can recover timing at the cost of area.
When we replace EPR with the grid or adjacency-list encoding, the resulting area degrades at every bit-width, indicating that global topology and node-level metadata (levels, fanout, parent/child links) help the LRM make better local edits.
To explain EPR's advantage, we evaluate how well the LRM reasons about a prefix graph under each representation. We construct three probes over randomly sampled prefix graphs, 100 questions each: fanout-cone size, node reachability, and node fanin-cone size. As reported in \Cref{table:repr_reasoning}, EPR attains at least 94\% accuracy on all three, while grid and adjacency-list fall to 58--63\% and 33--47\%. The LRM therefore recovers global structure far more reliably from EPR's explicit per-node attributes than from raw adjacency or grid positions alone, which accounts for the area gap in \Cref{table:ablation1}.
Similarly, omitting the critical-path prompts increases area at 32-bit and 64-bit, showing that highlighting the timing-critical region is beneficial.

To assess the contribution of CoT reasoning, we evaluate a non-reasoning variant of the LRM.
While the full PrefixAgent uses QwQ-32B model, the \emph{w/o CoT reasoning} setting in \Cref{table:ablation1} replaces it with its non-reasoning counterpart (Qwen2.5-32B), trained only with tool-call supervision and without CoT content.
This variant shows noticeably degraded QoR, confirming that CoT reasoning provides essential guidance beyond pure tool-call signals.

To assess the effect of training-data scale, we compare \textit{PrefixAgent} without fine-tuning against fine-tuning on 1/4 and 1/2 of the trajectories and on the full set, with the subsets sampled by trajectory count. QoR improves consistently as more training data is used, while the marginal gain diminishes as the dataset grows.

Finally, to assess robustness to the choice of LRM for CoT collection, we introduce an open-model variant in \Cref{table:cot_robust}.
Besides GPT-o3, we use the open-source Qwen3-235B-A22B-Thinking-2507 model to generate CoT content, keeping the prompt template and hyperparameters identical.
We obtain area–delay results and tool-call counts that closely match those of GPT-o3 across all evaluated bit-widths.
This indicates that PrefixAgent can operate effectively with different strong LRMs for CoT generation, supporting the robustness of the overall framework. Because the CoT teacher is the only proprietary component and can be replaced by the open-source Qwen3 model with comparable results, the pipeline is reproducible without proprietary models.

\subsection{Evaluation under a Commercial Flow and Implementation in Systolic Arrays}
\label{subsec:flow}
To evaluate \textit{PrefixAgent} under a different technology and implementation flow, we conduct experiments using a commercial physical implementation toolchain with a 32 nm technology.
We compare our method against adders synthesized by commercial synthesis tools (CST) using \texttt{y = a + b} style Verilog under non-uniform input arrival time profiles.
As shown in \Cref{table:commercial}, \textit{PrefixAgent} generates adders with a smaller area compared to CST solutions.

To assess benefits in an accelerator setting, we implement a 256-PE weight-stationary systolic array, where each PE uses two PrefixAgent-generated prefix adders: the final carry-propagate adder after the multiplier’s compressor tree and the accumulator adder.
As summarized in \Cref{table:array}, replacing the default synthesis-tool adders with PrefixAgent-generated adders reduces array-level area and also improves critical-path delay under both the OpenROAD flow and a commercial flow, for both 8-bit and 16-bit configurations.

\begin{table*}[!tb]
\caption{Comparison on commercial physical design flow}
\centering
\resizebox{\columnwidth}{!}{%
\begin{tabular}{|c|c|cc|cc|cc|}
\hline
\multirow{2}{*}{Input arrival profile} & \multirow{2}{*}{Method} & \multicolumn{2}{c|}{16-bit} & \multicolumn{2}{c|}{32-bit} & \multicolumn{2}{c|}{64-bit} \\
\cline{3-8}
& & Area ($\upmu$m$^2$) & Delay (ns) & Area ($\upmu$m$^2$) & Delay (ns) & Area ($\upmu$m$^2$) & Delay (ns) \\
\hline
\multirow{2}{*}{LSB-first}
& CST & 371.98 & 0.9995 & 728.38 & 1.0561 & 1639.09 & 1.7425 \\
& \textbf{PrefixAgent} & \textbf{349.19} & 0.9998 & \textbf{706.01} & 1.0490 & \textbf{1574.76} & 1.7453 \\
\hline
\multirow{2}{*}{Random}
& CST & 351.99 & 0.8781 & 697.37 & 1.2027 & 1767.82 & 1.5881 \\
& \textbf{PrefixAgent} & \textbf{333.75} & 0.8729 & \textbf{663.82} & 1.1944 & \textbf{1702.25} & 1.5863 \\
\hline
\end{tabular}
}
\label{table:commercial}
\end{table*}

\begin{table}[!tb]
\centering

\caption{Systolic array comparison under OpenROAD and commercial flows.
 (Area unit: $\upmu\text{m}^2$. Delay unit: ns).}

\begin{tabular}{|c|c|c|c|c|}
\hline
Flow & Bitwidth & Method & Area & Delay \\
\hline
\multirow{4}{*}{OpenROAD}
 & \multirow{2}{*}{8-bit}
   & Default     & 158069  & 0.9361 \\
 &                     & Ours & 156704  & 0.8716 \\
\cline{2-5}
 & \multirow{2}{*}{16-bit}
   & Default     & 465961  & 1.4640 \\
 &                     & Ours & 452625  & 1.3160 \\
\hline
\multirow{4}{*}{Commercial}
 & \multirow{2}{*}{8-bit}
   & Default     & 401701  & 1.7120 \\
 &                     & Ours & 391865  & 1.6987 \\
\cline{2-5}
 & \multirow{2}{*}{16-bit}
   & Default     & 1157562 & 2.3750 \\
 &                     & Ours & 1138543 & 2.3648 \\
\hline
\end{tabular}

\label{table:array}

\end{table}

\needspace{5\baselineskip}
\subsection{Comparison with Ling-Node and Physically-Aware Synthesis}
\label{subsec:ling}

\begin{table}[!tb]
\centering

\caption{Comparison with a physically-aware Ling-node baseline~\cite{Datapath-2008ASPDAC-Zhu} under the LSB-first non-uniform arrival profile (NanGate45, OpenROAD flow). Area in $\upmu$m$^2$, delay in ns.}
\label{table:ling}
\begin{tabular}{l cc cc cc}
\toprule
\multirow{2}{*}{Method} & \multicolumn{2}{c}{16-bit}
                        & \multicolumn{2}{c}{32-bit}
                        & \multicolumn{2}{c}{64-bit}\\
\cmidrule(lr){2-3}\cmidrule(lr){4-5}\cmidrule(lr){6-7}
 & Area & Delay & Area & Delay & Area & Delay\\
\midrule
Ling baseline~\cite{Datapath-2008ASPDAC-Zhu} & 247 & 0.6210 & 487 & 0.8095 & 1106 & 1.5184\\
PrefixAgent        & 217 & 0.5810 & 426 & 0.7980 &  938 & 1.4882\\
PrefixAgent + Ling & 226 & 0.5848 & 444 & 0.8033 &  971 & 1.4950\\
\bottomrule
\end{tabular}

\end{table}

To address physically-aware prefix co-optimization and Ling-node designs, we compare \textit{PrefixAgent} with a method~\cite{Datapath-2008ASPDAC-Zhu} that is both physically-aware and natively Ling-based, re-implemented in our flow under the LSB-first profile for a consistent comparison. As shown in \Cref{table:ling}, \textit{PrefixAgent} achieves smaller area than this baseline at every bit-width, up to 15.2\% at 64 bits, at equal or lower delay. Ling-node strategies are orthogonal to the prefix topology: a Ling adder shares the same prefix-processing network and only reformulates the carry recurrence, so any structure \textit{PrefixAgent} produces can be realized as a Ling adder. Applying Ling's recurrence to our structures (\textit{PrefixAgent}+Ling) increases area by about 4\% while leaving delay essentially unchanged, and \textit{PrefixAgent}+Ling still remains smaller than the baseline, indicating that the prefix structure rather than the Ling reformulation drives the improvement.

\subsection{Discussion}
\label{subsec:discussion}
In previous LLM-based methods, such as PrefixLLM~\cite{Datapath-2024Arxiv-Xiao}, the language model is tasked with generating the full prefix adder structure directly.
However, constructing a valid prefix adder graph requires maintaining intricate structural invariants, such as tree consistency, fan-in/fan-out correctness, and legality of prefix operations.
The model is expected to perform analysis, optimization, and validation simultaneously on the full graph, while the complexity of the task exceeds the LLM’s direct reasoning capabilities.
In contrast, \textit{PrefixAgent} decomposes the task into two modular phases: backbone generation and local refinement, allowing the LRM to focus its reasoning on manageable subproblems.
Equipped with domain-specific knowledge from pre-training and fine-tuning, the LRM benefits from the rich structural context provided by \textit{PrefixAgent}, enabling it to generalize its learned knowledge to novel designs.
Through this framework, the LRM is able to make high-level design decisions and invoke external tools to execute complex graph transformations, rather than attempting to manipulate the full structure directly.
Moreover, the use of high-quality optimization trajectories generated via e-graph rewrites enables effective supervised fine-tuning, further aligning the LRM’s reasoning process with domain-specific logic optimization profiles.
Search-based baseline methods require retraining for each new scenario and do not scale well with bitwidth due to the exponential design space.
As reported in the PrefixRL paper~\cite{RL-2021DAC-Roy}, optimizing a 64-bit adder requires over $5\times 10^5$ training steps, taking several weeks for one new scenario.
Moreover, we observe that these methods frequently explore high-deficiency structures that lead to excessive area consumption.
\textit{PrefixAgent} provides a flexible and efficient alternative across the evaluated design scenarios.

Empirically, we occasionally encounter pathological instances where, under very aggressive timing targets, further Phase~II refinements yield only marginal delay improvement at the cost of substantial area growth.
In such regimes, \textit{PrefixAgent} will backtrack to Phase~I again to propose a new backbone.

\section{Conclusion}
\label{sec:conclu}
We propose \textit{PrefixAgent}, a framework for prefix adder optimization. The design process is decomposed into two phases—backbone generation and local refinement.
To guide the LRM, we leverage high-quality optimization traces generated through e-graph rewrites.
Experimental results show that \textit{PrefixAgent} achieves smaller areas than the evaluated baselines in nearly all tested configurations, with the advantage growing at larger bit-widths.

Moreover, we introduce a novel paradigm for generating high-quality supervision traces, which remains valuable as LRM capabilities continue to advance.
Given recent e-graph–based optimizers for multipliers~\cite{10476812}, the same data-collection methodology can be reused to learn multiplier optimization policies.
In particular, compressor-tree templates (e.g., Wallace/Dadda trees) can be treated as multiplier backbones, and equivalence-preserving compressor rewrites can be used to generate trajectories for LRM fine-tuning.
Likewise, e-graph frameworks for general datapath RTL/netlist optimization~\cite{Egraph-2024TCAD-Coward,Egraph-2024DAC-Chen} indicate that this supervision paradigm can extend beyond prefix adders to broader datapath optimization tasks.

{
    \bibliographystyle{IEEEtran}
    \bibliography{ref/refs}
}

\end{document}